\documentclass{emulateapj}

\newcommand{\be}{\begin{equation}}
\newcommand{\ee}{\end{equation}}
\def\r{{\bf r}}
\def\k{{\bf k}}
\newcommand{\hompc}{\,h\,{\rm Mpc}^{-1}}
\newcommand{\mpcoh}{\,h^{-1}\,{\rm Mpc}}
\newcommand{\gs}{\mathrel{\lower0.6ex\hbox{$\buildrel {\textstyle >}
 \over {\scriptstyle \sim}$}}}
\newcommand{\ls}{\mathrel{\lower0.6ex\hbox{$\buildrel {\textstyle <}
 \over {\scriptstyle \sim}$}}}

\shorttitle{The SDSS DR5 Galaxy Power Spectrum}
\shortauthors{Percival et al.}

\begin{document}

\title{The shape of the SDSS DR5 galaxy power spectrum}

\author{
Will J.\ Percival\altaffilmark{1},
Robert C.\ Nichol\altaffilmark{1}, 
Daniel J.\ Eisenstein\altaffilmark{2},
Joshua A.\ Frieman\altaffilmark{3,4},
Masataka Fukugita\altaffilmark{5},
Jon Loveday\altaffilmark{6},
Adrian C.\ Pope\altaffilmark{7,8},
Donald P.\ Schneider\altaffilmark{9},
Alex S.\ Szalay\altaffilmark{7},
Max Tegmark\altaffilmark{10},
Michael S. Vogeley\altaffilmark{11},
David H.\ Weinberg\altaffilmark{12},
Idit Zehavi\altaffilmark{13},
Neta A.\ Bahcall\altaffilmark{14},
Jon Brinkmann\altaffilmark{15},
Andrew J.\ Connolly\altaffilmark{16},
Avery Meiksin\altaffilmark{17},
}

\begin{abstract}

  We present a Fourier analysis of the clustering of galaxies in the
  combined Main galaxy and Luminous Red Galaxy (LRG) Sloan Digital Sky
  Survey (SDSS) Data Release 5 (DR5) sample. The aim of our analysis
  is to consider how well we can measure the cosmological matter
  density using the signature of the horizon at matter-radiation
  equality embedded in the large-scale power spectrum. The new data
  constrains the power spectrum on scales $100$--$600\mpcoh$ with
  significantly higher precision than previous analyses of just the
  SDSS Main galaxies, due to our larger sample and the inclusion of
  the LRGs. This improvement means that we can now reveal a
  discrepancy between the shape of the measured power and linear CDM
  models on scales $0.01<k<0.15\hompc$, with linear model fits
  favouring a lower matter density ($\Omega_M=0.22\pm0.04$) on scales
  $0.01<k<0.06\hompc$ and a higher matter density
  ($\Omega_M=0.32\pm0.01$) when smaller scales are included, assuming
  a flat $\Lambda$CDM model with $h=0.73$ and $n_s=0.96$. The lower
  matter density favoured by fitting our SDSS data for
  $0.01<k<0.06\hompc$ is a better match to the best-fit WMAP 3-year
  cosmological model, and to results from the positions of the baryon
  oscillations observed in the SDSS DR5 power spectrum. This
  discrepancy could be explained by scale-dependent bias and, by
  analysing subsamples of galaxies, we find that the ratio of
  small-scale to large-scale power increases with galaxy luminosity,
  so all of the SDSS galaxies cannot trace the same power spectrum
  shape over $0.01<k<0.2\hompc$. However, the data are insufficient to
  clearly show a luminosity-dependent change in the largest scale at
  which a significant increase in clustering is observed, although
  they do not rule out such an effect. Significant scale-dependent
  galaxy bias on large-scales, which changes with the $r$-band
  luminosity of the galaxies, could potentially explain differences in
  our $\Omega_M$ estimates and differences previously observed between
  2dFGRS and SDSS power spectra and the resulting parameter
  constraints.

\end{abstract}


\altaffiltext{1}{Institute of Cosmology and Gravitation, Mercantile
  House, Hampshire Terrace, University of Portsmouth, Portsmouth, P01
  2EG, UK}
\altaffiltext{2}{Steward Observatory, University of Arizona, 933
  N. Cherry Ave., Tucson, AZ 85121, USA}
\altaffiltext{3}{Particle Astrophysics Center, Fermilab, P.O. Box 500, 
  Batavia, IL 60510, USA}
\altaffiltext{4}{Kavli Institute for Cosmological Physics, 
  Department of Astronomy \& Astrophysics, University of Chicago, 
  Chicago, IL 60637, USA}
\altaffiltext{5}{Institute for Cosmic Ray Research, University of Tokyo, 
  Kashiwa 277-8582, Japan}
\altaffiltext{6}{Astronomy Centre, University of Sussex, Falmer, Brighton, 
  BN1 9QH, UK}
\altaffiltext{7}{Department of Physics and Astronomy,
  The Johns Hopkins University,
  3701 San Martin Drive, Baltimore, MD 21218, USA}
\altaffiltext{8}{Institute for Astronomy, University of Hawaii, 
    2680 Woodlawn road, Honolulu, HI 96822, USA}
\altaffiltext{9}{Department of Astronomy and Astrophysics, The 
  Pennsylvania State University, University Park, PA 16802, USA}
\altaffiltext{10}{Department of Physics, Massachusetts Institute of
  Technology, Cambridge, MA 02139, USA}
\altaffiltext{11}{Department of Physics, Drexel University, Philadelphia, 
  PA 19104, USA}
\altaffiltext{12}{Department of Astronomy, The Ohio State University,
  Columbus, OH 43210, USA}
\altaffiltext{13}{Department of Astronomy, Case Western Reserve University, 
  Cleveland, OH 44106, USA}
\altaffiltext{14}{Department of Astrophysical Sciences, Princeton University,
  Princeton, NJ 08544, USA}
\altaffiltext{15}{Apache Point Observatory, P.O. Box 59, Sunspot, NM 88349, USA}
\altaffiltext{16}{Department of Physics and Astronomy, University of 
    Pittsburgh, Pittsburgh, PA 15260, USA}
\altaffiltext{17}{SUPA; Institute for Astronomy, University of Edinburgh,
  Royal Observatory, Blackford Hill, Edinburgh, EH9 3HJ, UK}

\section{INTRODUCTION}

The evolution of perturbations in the early universe imprints
characteristic scales that depend on the average matter density
\citep{silk68,peebles70,sunyaev70,bond84,bond87,holtzman89}.
Fundamentally, the growth of fluctuations is intimately linked to the
Jeans scale; perturbations smaller than the Jeans scale do not
collapse due to pressure support, while larger perturbations are free
to grow through gravity. In the radiation dominated era, the dark
matter has negligible density compared to the photon-baryon fluid, and
the perturbations in this fluid are stabilised by the high radiation
pressure at a time when the sound speed was of order
$c/\sqrt{3}$. Consequently, the Jeans scale is of order the horizon
scale until matter-radiation equality, after which it reduces to zero
when the matter dominates. We therefore see that the horizon scale at
matter-radiation equality will be imprinted in the distribution of
fluctuations -- this scale marks a turn-over in the growth rate of
fluctuations.

In a model with only collisionless dark matter, all lengths scale with
the horizon scale at matter radiation equality, which is a multiple of
$(\Omega_Mh^2)^{-1}$, where $h=H_0/100\,{\rm
  km\,s^{-1}Mpc^{-1}}$. Consequently, CDM transfer function fitting
formulae (such as Eq. G3 of \citealt{bbks}) were traditionally created
as a function of $q\equiv k / (\Omega_Mh^2 {\rm Mpc})$. When analysing
galaxy redshift surveys, the comoving distance--redshift relation
introduces another factor of $h$ into the measurements, so the data
constrain the transfer function in $ k / \hompc$. Consequently, fits
between model and data constrain the degenerate parameter combination
$\Omega_Mh$.

Such cosmological constraints are important for breaking degeneracies
between cosmological parameters that exist when fitting just CMB data
\citep{eisenstein99,efstathiou99,percival02}. With increasingly
precise temperature and polarisation CMB constraints
\citep{hinshaw06,page06,spergel06}, the additional cosmological role
of galaxy surveys to provide cross-checks is becoming increasingly
important. In this paper, we carefully test the methodology behind
using the shape of the galaxy power spectrum to provide cosmological
constraints.

Extracting the cosmological information encoded in the galaxy power
spectrum has motivated many previous surveys. Early studies, with of
order $10^4$ galaxies, include the CfA \citep{vogeley92,park94}, LCRS
\citep{shectman96} and PSCz \citep{saunders00} surveys. These surveys
were able to show that the shape of the power spectrum required a
relatively small matter density, at odds with the simple Einstein-de
Sitter model (e.g. \citealt{efstathiou90,tadros99}). The same general
shape of the power spectrum has been recovered by deprojecting the APM
\citep{maddox90,maddox96} survey, the parent catalogue of the 2dFGRS
\citep{efstathiou01,PB03}, and by deprojecting the photometric
component of the SDSS catalogue
\citep{scranton02,connolly02,tegmark02,dodelson02,szalay03}. More
recently, photometric redshifts have been exploited to improve the
deprojection of galaxy distances and the accuracy with which the power
spectrum can be measured from photometric data. Two studies have
recently been completed considering Luminous Red Galaxies (LRGs)
within the SDSS survey \citep{padmanabhan06,blake06}, showing a marked
improvement in accuracy on previous work.

However, the most precise measurement of the 3-dimensional galaxy
power spectrum is still recovered from redshift surveys. Over the past
5 years there has been a revolution in the number of galaxy redshifts
measured, and the size of the volume probed by these surveys. Leading
the way have been two large galaxy redshift surveys, the 2dF Galaxy
Redshift Survey (2dFGRS; \citealt{colless01,colless03}) and Sloan
Digital Sky Survey (SDSS; \citealt{york00}). In this paper we analyse
the relative clustering strengths of galaxy samples as a function of
scale. The overall clustering amplitude is also potentially
interesting in constraining cosmological models, but takes more work
to decode. This normalisation is known to be a strong function of both
galaxy colour and luminosity
\citep{park94,norberg01,norberg02,zehavi02}, so an understanding of
galaxy bias is required before cosmological constraints can be derived
from this statistic. However, the effect of galaxy bias on the shape
of the power spectrum is less certain, and is the subject of this
paper. On small scales, where non-linear corrections to the matter
power spectrum are important, the shapes of galaxy power spectra are
known to depend on galaxy colour and luminosity
(e.g. \citealt{cole05}). However, on large-scales, the effects on the
shape are more uncertain.

On larger scales where the matter in the universe is still expected to
be predominantly in the linear regime $k\ls0.15\hompc$
\citep{smith03}, discrepancies currently exist between the shapes of
the power spectra recovered from the 2dFGRS and SDSS. The problem is
demonstrated by the measurements of $\Omega_M$ obtained from such
fits. Assuming a Hubble parameter $h=0.72$, the \citet{tegmark04} SDSS
main galaxy analysis favoured $\Omega_M=0.296\pm0.032$. Similar values
of $\Omega_M\simeq0.3$ were found by alternative analyses of the red
selected SDSS main galaxy power spectrum \citep{pope04}, a set of SDSS
spectroscopic Luminous Red Galaxies (LRGs) \citep{eisenstein05}, and
SDSS photometrically selected LRGs \citep{padmanabhan06,blake06}. In
combination with the WMAP 1-year data \citep{spergel03}, the
\citet{eisenstein05} analysis of the LRGs provided the constraint
$\Omega_M=0.273\pm0.025$ from a combination of the overall shape of
the correlation function and the peak caused by baryon acoustic
oscillations. In contrast, a lower matter density
$\Omega_M=0.231\pm0.021$ is favoured by measurements of clustering of
blue selected galaxies in the 2dFGRS \citep{cole05}, who included a
simple model for scale-dependent bias, although this had a relatively
minor effect on the recovered matter density. Recent results from the
3-year WMAP data have provided an independent constraint on the matter
density, finding $\Omega_M=0.234\pm0.035$ (mean constraint from Table
2 of \citealt{spergel06}) from a better resolution of the third peak
height. It is clear that these discrepancies between measurements are
not linked to a single technique or particular analysis.

When the constraints from the galaxy power spectra are combined with
the CMB data the discrepancy is still clear. From Table~5 of
\citet{spergel06}, we see that combining the 3-year WMAP data with
additional cosmological constraints from the 2dFGRS power spectrum of
\citet{cole05} gives $\Omega_M=0.236^{+0.016}_{-0.029}$. However, when
the WMAP data is combined with constraints from the SDSS power
spectrum of \citet{tegmark04}, the higher $\Omega_Mh$ constraint from
the large-scale structure data increases the best fit to
$\Omega_M=0.266^{+0.025}_{-0.040}$ (see Table~6 of
\citealt{spergel06}).

The tension between measurements from different large-scale structure
experiments and the CMB observations is at the level of approximately
$2\sigma$, and it is therefore possible that it could be explained by
cosmic variance. Alternative explanations include a scale-dependent
galaxy bias on scales $k\ls0.15\hompc$, or a systematic problem with
one of the data sets. In this paper we test these possibilities by
measuring the redshift-space power spectrum from the latest SDSS
sample, Data Release 5 (DR5), which contains approximately twice as
many galaxies as previously analysed, and 60\% more LRGs than used in
\citet{eisenstein05}. By optimally analysing all of the galaxies to
calculate the underlying power spectrum, we obtain the most accurate
determination of the redshift-space power spectrum ever obtained for
any sample of galaxies. First, we will test if the discrepancy between
previous 2dFGRS and SDSS power spectra remains. Second, the size of
the sample means that we can consider how well the galaxies trace the
mass, by testing the simple hypothesis that the shape of the power
spectrum matches linear CDM models over $k<0.15\hompc$, and looking
for discrepancies over these scales. Finally, the number of galaxies
and volume covered are now sufficient to accurately measure power
spectra for subsamples, splitting the catalogue to test for general
changes in the shape of the power spectrum as a function of galaxy
properties.

In a parallel paper, \citet{tegmark06} present an analysis of a
largely overlapping data set, drawn from SDSS DR4, with a focus on the
implications for multi-parameter cosmological model fits. There are a
number of differences in the analysis methods. \citet{tegmark06} use a
pseudo-Karhoenen-Loeve method \citep{vogeley96,tegmark97} to estimate
the real space galaxy power spectrum, using finger-of-god compression
and linear theory to remove redshift-space distortion effects.  We use
the Fourier method of \citet{PVP}, which extends that of \citet{FKP},
to estimate the angle-averaged (monopole) redshift-space galaxy power
spectrum.  We combine the LRG and main galaxy samples, while
\citet{tegmark06} concentrate on the LRGs, after showing that they
have a power spectrum shape consistent with that of the main sample.
In addition, the many technical decisions that go into these analyses,
regarding completeness corrections, angular masks, K-corrections, and
so forth, were made independently for the two papers, and they present
different tests for systematic uncertainties.  Despite these many
differences of detail, our conclusions agree to the extent that they
overlap (this will be discussed in \citealt{tegmark06}), a reassuring
indication of the robustness of the results.

\section{THE SDSS DR5 SAMPLE}  \label{sec:dr5}

The Sloan Digital Sky Survey (SDSS;
\citealt{york00,stoughton02,abazajian03,abazajian04,adelman06a}) is an
ongoing survey using a 2.5m telescope \citep{gunn06} to obtain $10^4$
square degrees of imaging data in five passbands $u$, $g$, $r$, $i$
and $z$ \citep{fukugita96,gunn98}. The images are reduced
\citep{lupton01,stoughton02,pier03,ivezic04} and calibrated
\citep{lupton99,hogg01,smith02,tucker06}, and galaxies are selected in
two ways for follow-up spectroscopy. The main galaxy sample
\citep{strauss02} targets galaxies brighter than $r=17.77$
(approximately $90$ per square degree, with a median redshift
$z=0.11$); in this paper we use the DR5 sample \citep{adelman06b}
containing 465789 main galaxies that meet our selection criteria. In a
small subset of the data taken during initial survey operation, we set
a conservative faint magnitude limit corresponding to $r=17.5$, to
avoid minor fluctuations in the survey depth.

In addition to the main galaxy sample, we also select 56491 cut-I and
cut-II Luminous Red Galaxies (LRGs; \citealt{eisenstein01}). The
selection of these galaxies, based on $g$, $r$ and $i$ colours and
going to a deeper $r$ magnitude, adds approximately $15$ galaxies per
square degree in addition to the main sample, and extends the redshift
distribution to $z\simeq0.5$. In our sample, 65032 of the main
galaxies are also targeted in the SDSS as LRGs, but only 21310 of
these galaxies are intrinsically luminous with ${\rm
  M}_{^{0.1}r}<-21.8$ (${\rm M}_{^{0.1}r}$ is defined in
Section~\ref{sec:gal_mags}). Our sample therefore consists of 77801
bright LRGs and 444479 other galaxies, giving 522280 galaxies in
total.

These galaxies and objects selected for SDSS spectroscopic observation
for other scientific programmes are assigned to plug-plates using a
tiling algorithm designed to ensure nearly complete samples
\citep{blanton03a}. The spectra are good enough to allow redshifts to
be obtained for almost all of the galaxies selected for observation. A
detailed review of the SDSS galaxy observing strategy and the main
galaxy sample is given by \citet{tegmark04}.

In our investigation, we simultaneously analyse the main galaxy sample
and the LRGs, therefore including correlations between the two data
sets in addition to internal correlations within the individual
subsamples. This combination is aided by the fact that the transition
from main galaxies to LRGs within the survey is smooth in terms of
galaxy properties and expected bias. In this section we present an
overview of the data and the techniques used to model the selection
function of this combined sample.

Our chosen analysis method uses an empirically determined model of
luminosity-dependent (but scale-independent) bias to increase the
accuracy with which the underlying power spectrum can be recovered,
and correct for offsets in the measured power caused by such a bias (a
description of the model is provided in Section~\ref{sec:bias}). We
apply two cuts to the final DR5 sub-sample that we analyse as a
consequence of this model: we exclude low luminosity LRGs with ${\rm
  M}_{^{0.1}r}>-21.8$, and high luminosity galaxies ${\rm
  M}_{^{0.1}r}<-23.0$ (whether main galaxy or LRG). The reasons for
these cuts are presented in Section~\ref{sec:bias}.

Due to practical limitations of fibre positioning, spectra cannot be
obtained for objects closer than 55\,arcsec, within a single
spectroscopic tile. This is mitigated to some extent by multiple
observations of the same region where tiles overlap, but we choose to
apply a further correction; if a targeted main galaxy or LRG has no
redshift and lies within 55\,arcsec of another main galaxy or LRG with
a redshift, then the observed redshift is assigned to both
galaxies. Main galaxies (non-LRGs) and LRGs are treated separately;
where a main galaxy or LRG is not assigned a fibre due to collision
with a quasar or galaxy that is not in the same class (e.g. a main
galaxy is obscured by a LRG), then no redshift is assigned. This is
the procedure adopted by \citet{zehavi02} where, by comparison with
just the plate overlap regions where fibre collision has a reduced
impact, it was shown to provide a sufficient correction for this
effect on the large-scales of interest in the current study. We apply
a close-pair correction to $3132$ LRGs and $19402$ main galaxies,
which form 4.3\% of the total population.

In certain regions observed early in the survey operations, we exclude
all main galaxies with $r>17.5$, so the main galaxies form two
populations with $r<17.5$, or $r<17.77$ depending on angular
position. 49688 of the main sample galaxies (11.9\%) lie in the
regions limited to $r<17.5$. We assume that the LRGs form a single
population with an isotropic redshift distribution. The radial
selection function of our combined sample can therefore be decomposed
into three sub-samples with different radial distributions -- main
galaxies with $r<17.5$, main galaxies with $r<17.77$, and LRGs. The
limits of the survey are trimmed by setting redshift limits on the
combined sample by only considering galaxies with $0.003<z<0.5$. LRGs
not in the main galaxy sample must have $z>0.15$, and main sample
galaxies must have $z<0.3$. This removes regions where the selection
function becomes small and poorly determined.

\begin{figure}[tb]
  \plotone{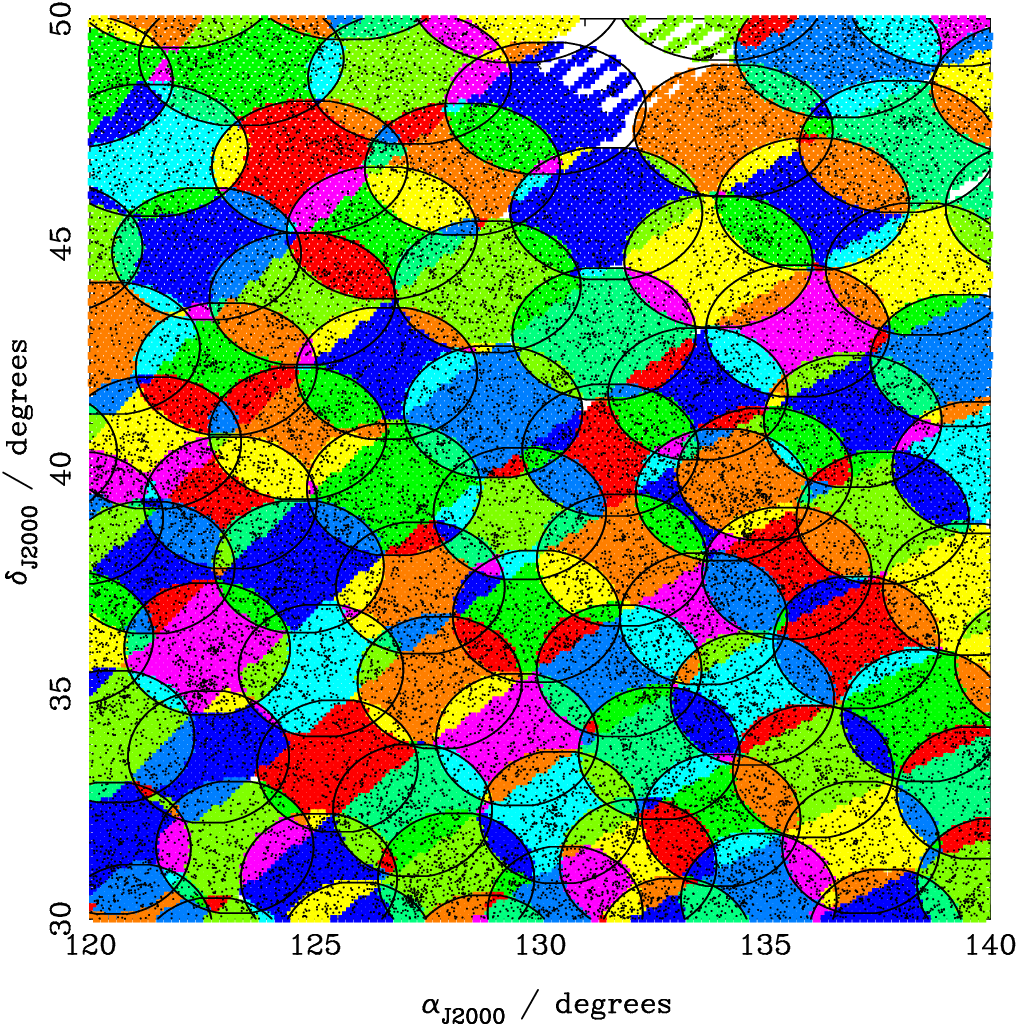}
  \caption{Part of the SDSS DR5 sample region plotted as a function of
    celestial coordinates. Note that this is not an equal area
    projection, so the plate outlines (solid lines) do not form
    perfect circles. Different colours delineate groups of pixels
    where the completeness of the survey is expected to be the same
    (see text for details). Black points show the positions of the
    galaxies. White regions are bad areas excluded from the survey
    mask. Where these areas are due to bad photometric fields, the
    regions often follow the drift scanning strategy of the
    photometric survey -- hence the white stripes at
    $\alpha\simeq132^{\circ}$,
    $\delta\simeq48^{\circ}$. \label{fig:ra_dec_region}}
\end{figure}

\subsection{Angular Selection} \label{sec:sel_ang}

The angular selection function of the SDSS galaxies was modelled using
a routine based on a HEALPIX \citep{gorski05} equal-area
pixellization of the sphere. HEALPIX was used to decompose the sphere
into $3145728$ pixels, each of size 0.013\,deg$^2$. This means that
each SDSS plate is covered by 532 pixels, and the DR5 sample analysed
covers $487177$ pixels. Given the large angular scales of interest,
the effect of this pixellization should be negligible on the resulting
power spectrum.

The first task is to find groups of pixels that have the same
spectroscopic targeting information -- they cover regions selected in
the same targeting run(s), covered by the same set of tiles, and are
within the photometric SDSS survey region -- i.e. not in a bad
field. The SDSS photometric survey area is decomposed into fields of
approximate size $0.033$ square degrees. Consequently, bad fields, as
defined in the SDSS Catalogue Archive, often only cover part of a
pixel, so for each pixel we have allowed the effective area to be
reduced by bad fields rather than removing the whole pixel. There are
small internal regions within the area covered by the SDSS imaging
that are not covered by spectroscopic tiles, which only have a few
possible target galaxies, if any at all. We have included such regions
as separate groups of pixels in our analysis. In total, we decompose
the survey region in 6447 distinct groups. Note that we use a full
list of targeted plates to create this decomposition of the spherical
surface rather than just the observed plates; it is the targeting
algorithm \citep{blanton03a} that decides the distribution of the
spectroscopically observed galaxies, so the distribution across
observed plates will depend on the positions of the unobserved
overlapping plates.

An example region within the angular mask created for the survey is
plotted in Fig.~\ref{fig:ra_dec_region}, showing the different
groups. The staves (named for the similarly shaped planks in barrel
making) within the survey, corresponding to photometric sections from
different great circle scans, are clearly visible. The most obvious
pattern in the decomposition of the angular mask is due to the overlap
of different spectroscopic plates, which can leave some small regions
containing only a handful of galaxies. The boundaries to different
targeting regions follow segments of staves, leading to additional
divisions across different staves. This region was chosen as it
includes a small area covered by bad fields that shows the interleaved
drift-scan strategy of the SDSS photometric observations, splitting
the stave into 12 individual columns.

Having created a list of target regions on our pixellated mask, we
calculated the completeness within each group, the ratio of good
quality spectra to targets. Any small internal group containing no
targets was assigned a completeness of 1. We have applied a
completeness cut of 70\% and exclude regions with a lower
completeness. In general this only removes regions where the
spectroscopic survey is incomplete and further observations are
scheduled for a particular group. For our final sample, 97.6\% of the
galaxies targeted have redshifts with spectroscopic confidence greater
than 80\%, as defined in the SDSS Catalogue Archive Server (CAS),
after the fiber collision correction mentioned above.

\begin{figure}[tb]
  \plotone{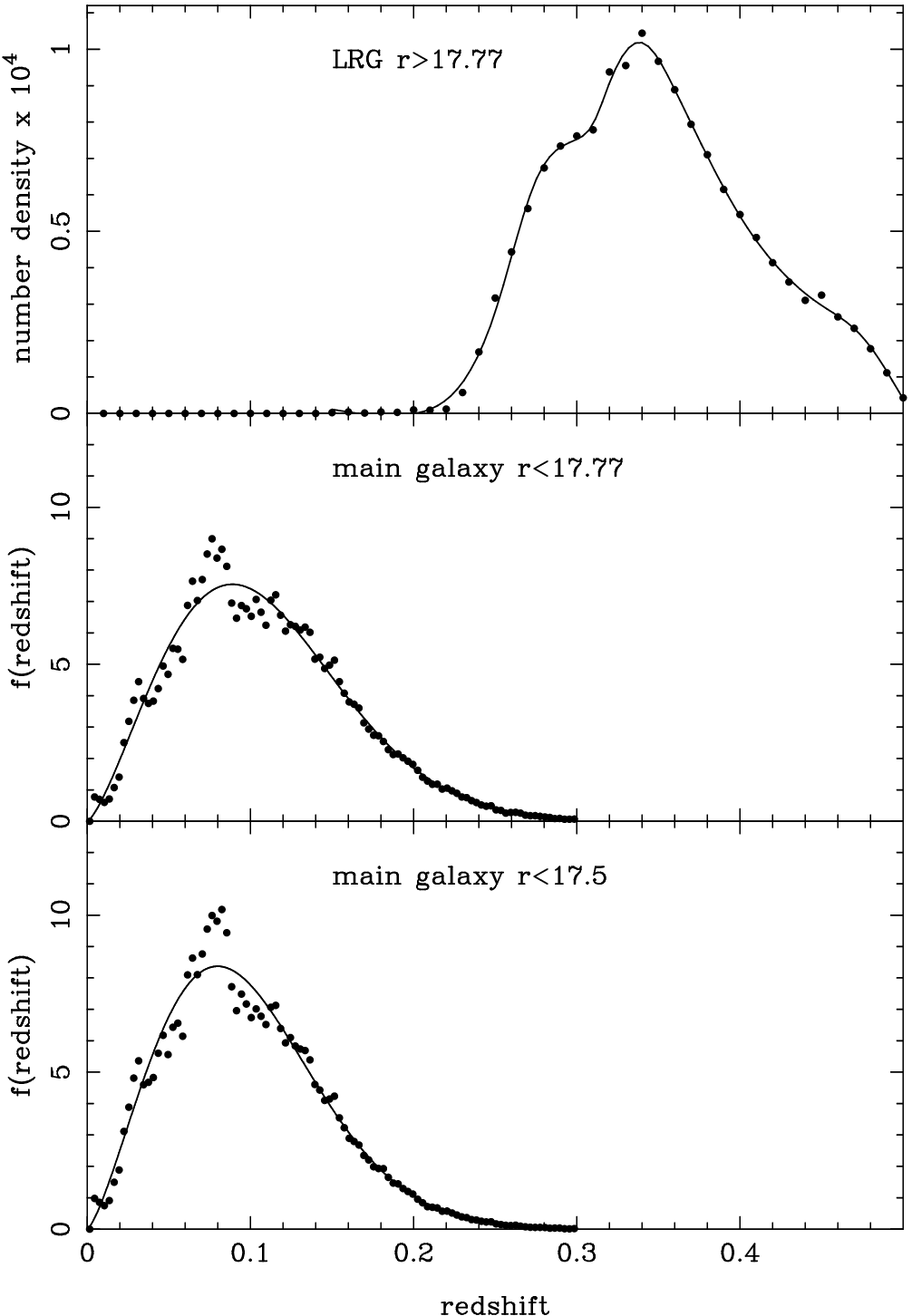}
  \caption{Radial distributions of the SDSS DR5 main galaxies (solid
    circles) to an apparent r magnitude limit of 17.5 (lower panel),
    and 17.77 (middle panel). The sharp increase in the number of
    galaxies at $z=0.08$ is predominantly the effect of the ``Sloan
    Great Wall'' \citep{gott05}. These data are fitted using equation
    (\ref{eq:zfit}), shown by the solid lines \citep{baugh93}. In the
    upper panel we show the distribution of the number density of the
    LRGs (excluding those in the main galaxy sample). These data are
    fitted with a smooth cubic spline (solid lines) with nodes
    selected to allow the curve to fit the sharp distortions caused by
    spectroscopic features moving through the SDSS filters used to
    select the LRGs, and the mixture of cut-I and cut-II LRGs (see
    text for details). \label{fig:red_dist}}
\end{figure}

\subsection{Radial Selection} \label{sec:sel_rad}

In order to fully quantify the expected distribution of galaxies, we
obviously need to model the radial galaxy distribution. We do this in
different ways for the main galaxies and LRGs. For the main galaxies,
we have found that a simple fit of the form \citep{baugh93}
\be
  f(z) = z^g \exp\left[-\left(\frac{z}{z_c}\right)^b\right],
  \label{eq:zfit}
\ee
provides an adequate fit to the data, where $g$, $b$ and $z_c$ are
fitted parameters. Fig.~\ref{fig:red_dist} shows the distribution of
main galaxy redshifts in the SDSS DR5 sample compared with a fit of
this form for apparent magnitude limits of $r=17.5$ and $r=17.77$; as
discussed in Section~\ref{sec:sel_ang}, in some angular directions the
effective magnitude limit had to be reduced to $r=17.5$. For galaxies
with $r<17.5$, the best-fit model redshift distribution has
$z_c=0.0955$, $b=1.88$, $g=1.35$. For galaxies with $r<17.77$, the
best-fit value of $z_c$ changes to $z_c=0.106$.

For the LRGs, we fit the radial number density using a cubic spline
\citep{press92} with nodes separated by $\Delta z=0.05$, although we
add additional nodes to enable the fit to match the distortions at
$z\simeq 0.3,\, 0.34,\, 0.44$, caused by spectroscopic features moving
through the SDSS filters used to select the LRGs, and the join between
cut-I and cut-II LRGs. At the higher redshifts probed by the LRGs, the
effect of clustering on the redshift distribution diminishes, and a
spline fit is less likely to remove structure compared with lower
redshift data.

\subsection{Luminosities} \label{sec:gal_mags}

Where specified, we have K-corrected the galaxy luminosities using the
methodology outlined in \citet{blanton03a,blanton03b}. In particular,
we have used the {\tt kcorrect\_v4\_1\_4} software package using the
observed $u$,$g$,$r$,$i$,$z$ Galactic extinction-corrected (using the
maps of \citealt{schlegel98}) Petrosian magnitudes from the DR5 CAS
archive and their measured errors. We have used the ``BEST'' database
to obtain the galaxy magnitudes, and the ``TARGET'' database to obtain
the list of galaxy targets. We also use the same $z=0.1$ shifted
$r$-band filter to define our luminosities (as discussed in
\citealt{blanton03b}), which we refer to as M$_{^{0.1}r}$ throughout
this paper. Galaxy luminosities without K-corrections are written as
M$_r$.  We use the same method for both the LRG sample and the main
SDSS galaxy sample, and have not added evolutionary corrections. We
have assumed $\Omega_M=0.3$ and $\Omega_\Lambda=0.7$ for computing
these K-corrections and applied recommended AB corrections to the
observed SDSS magnitude system \citep{smith02}.

\subsection{Colours}

\begin{figure}[tb]
  \plotone{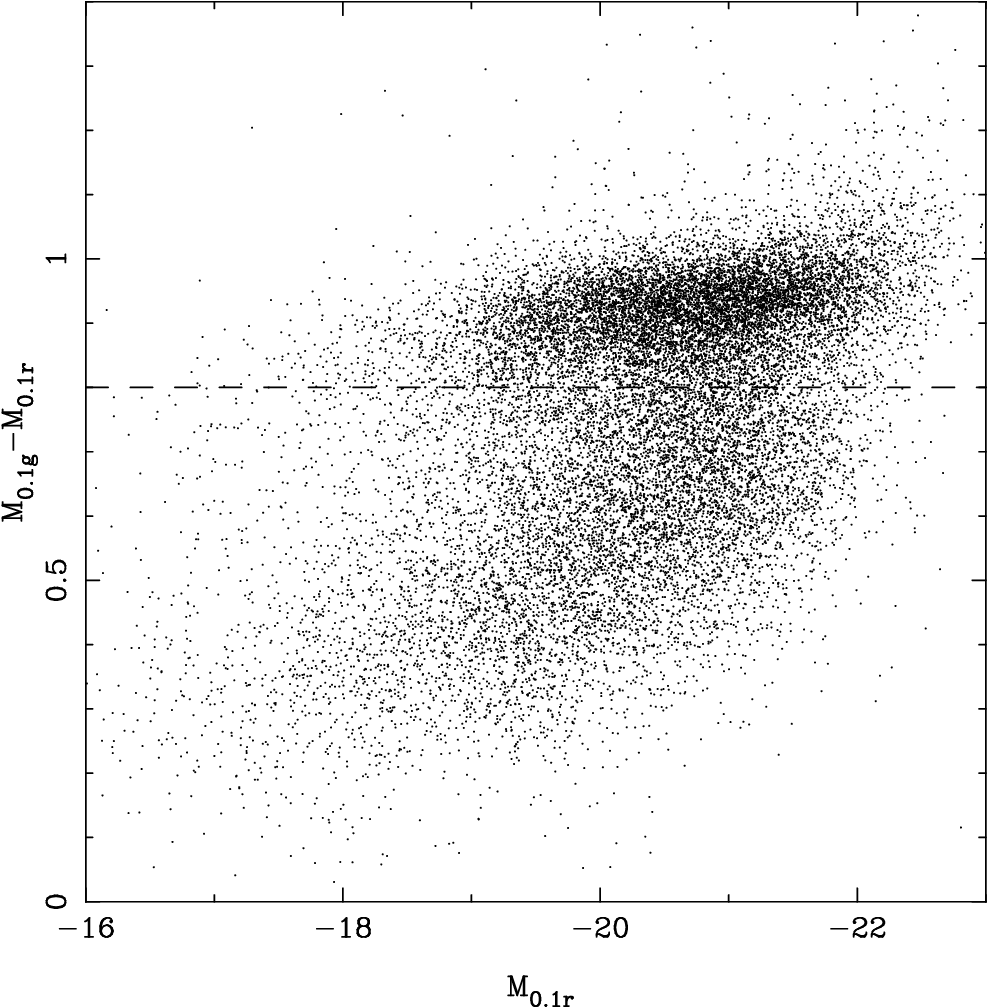}
  \caption{The colour of 5\% of the main galaxies in the SDSS DR5
    sample, selected at random. The well-known bimodal split between
    red and blue galaxies is clear, approximately split by ${\rm
      M}_{^{0.1}g}-{\rm M}_{^{0.1}r}=0.8$ (dashed line). This plot
    highlights the fact that as we change the magnitude of the
    samples, we also change the average colour, with the more luminous
    main galaxies being predominantly redder. \label{fig:gmr_vs_r}}
\end{figure}

The average colours of galaxies change as the luminosity
increases. Fig.~\ref{fig:gmr_vs_r} shows the ${\rm M}_{^{0.1}g}-{\rm
  M}_{^{0.1}r}$ colour distribution of these galaxies, plotted as a
function of ${\rm M}_{^{0.1}r}$. The well-known bimodal red--blue
split in galaxy colours \citep{strateva01,baldry04} is evident, and
can be approximately delineated by ${\rm M}_{^{0.1}g}-{\rm
  M}_{^{0.1}r}=0.8$.  Importantly, for our development of a bias model
for these galaxies, as we change the magnitude we also change the
colour of our sample, with more luminous galaxies being predominantly
redder. We will see in the next section that this means that there is
a smooth transition between the main galaxy and LRG samples.

\begin{figure}[tb]
  \plotone{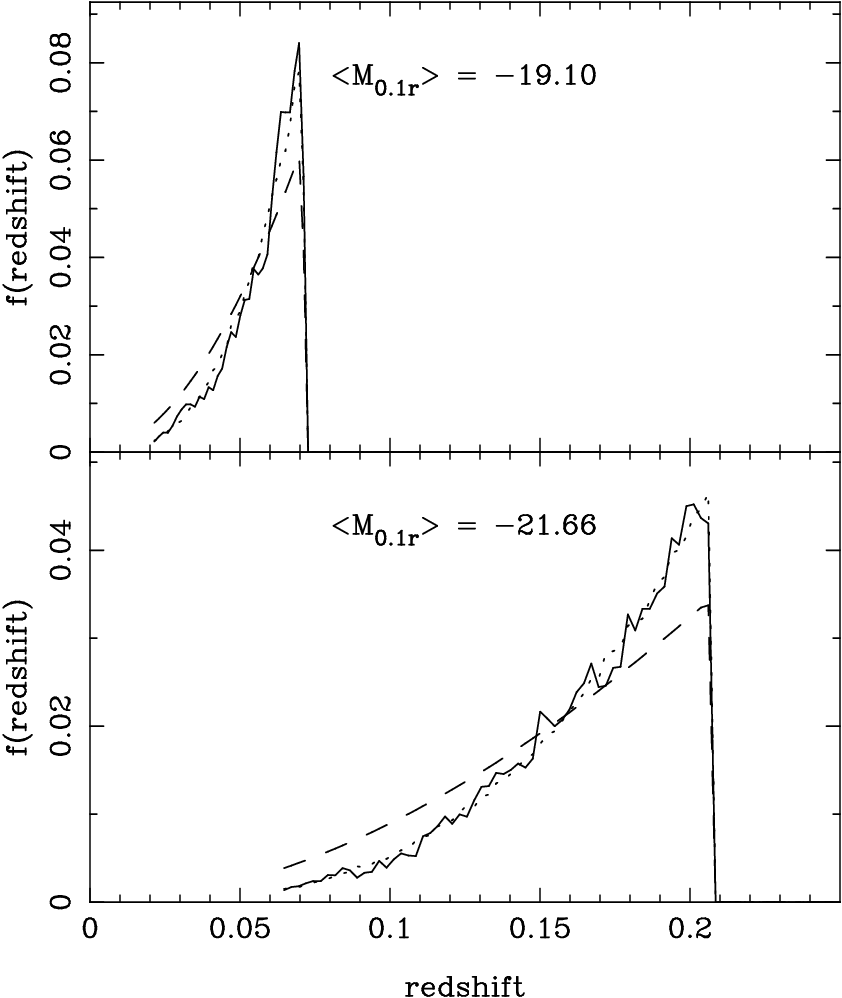}
  \caption{The redshift distributions of two of our pseudo-volume
    limited catalogues (solid lines). The selection of these
    catalogues was based on absolute magnitudes without K-corrections
    or corrections for evolution, and so are not strictly volume
    limited. These effects were instead included in estimating the
    redshift distribution of the samples to create matched random
    catalogues (dotted lines; see text for details). For comparison,
    the dashed curves show the redshift distribution that the catalogues
    would have if they were strictly volume limited -- as can be seen
    these are a poor fit to the data. \label{fig:absmag_cat_zdist}}
\end{figure}

\section{MODELLING SCALE-INDEPENDENT GALAXY BIAS}  \label{sec:bias}

For a magnitude limited survey, on average, the largest scales will be
probed by the most luminous galaxies. Luminous galaxies are known to
be more biased than less-luminous galaxies (e.g. \citealt{park94}), so
this relative bias needs to be quantified in order to minimise any
systematic effects caused by comparing different types of galaxies on
different scales. For the main galaxy sample we analyse subsamples
with different average luminosity, in order to measure the relative
bias (Section~\ref{sec:bias_maingal}). The LRG sample is now
sufficiently large that we can also consider splitting this catalogue
into subsamples as a function of luminosity (this is discussed in
Section~\ref{sec:bias_lrg}).

\subsection{Main Galaxies}  \label{sec:bias_maingal}

\begin{figure}[tb]
  \plotone{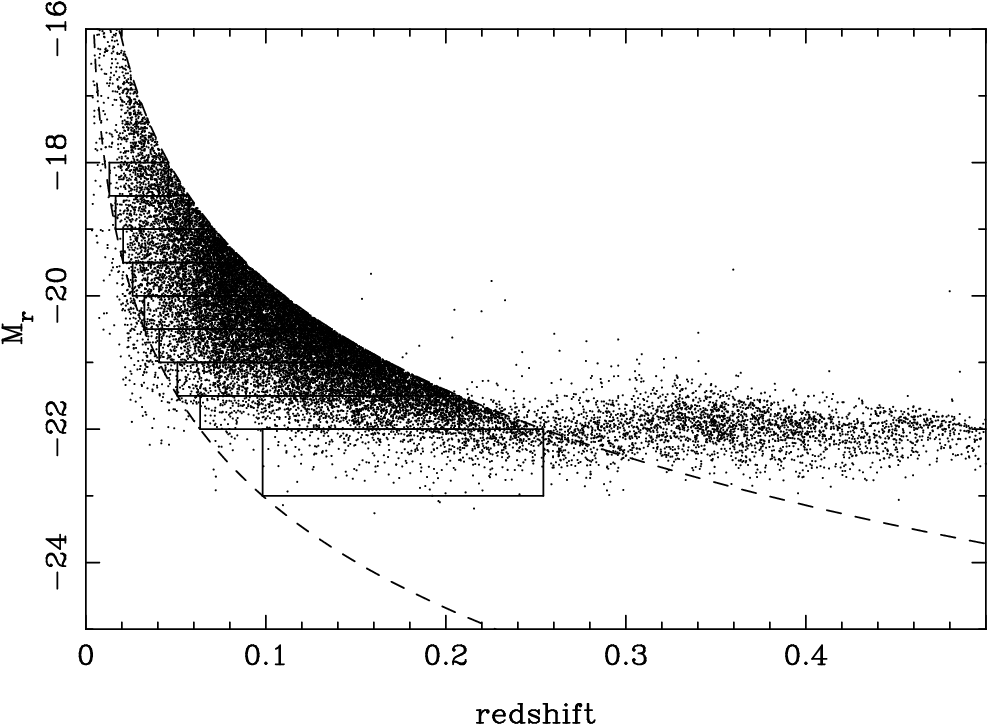}
  \caption{The distribution of the SDSS galaxies in the redshift --
    luminosity plane. Absolute magnitudes were calculated assuming a
    flat $\Lambda$ cosmology with $\Omega_M=0.3$ and have not been
    K-corrected or corrected for evolution. The upper and lower
    apparent magnitude limits of the main galaxy sample are shown by
    the dashed lines. The redshift and magnitude limits of the
    pseudo-volume limited catalogues analysed to calculate the
    relative bias as a function of absolute magnitude are shown by the
    overlaid rectangles. \label{fig:z_vs_R}}
\end{figure}

In order to empirically determine the relative clustering strengths of
the main galaxy sample as a function of luminosity, we split the
catalogue based on the absolute magnitude without K-correction or
correction for evolution in the luminosity function, and the
corresponding redshift limits where the sub-catalogue is complete (see
Fig.~\ref{fig:z_vs_R}). We analyse 8 catalogues of width $\Delta{\rm
  M}_r=0.5$ with $-22.0\le{\rm M}_r\le-18.0$, and an additional
catalogue of bright galaxies with $-23<{\rm M}_r<-22$. We call these
catalogues pseudo-volume limited, as they are not strictly volume
limited because we have ignored K-corrections and evolution in the
luminosity function. In order to estimate the redshift distribution of
each sample, we determine the average K-correction as a function of
redshift, and use this, together with the best-fit evolutionary
corrections of \citet{blanton03b}, to predict the expected galaxy
number density. The redshift distributions for two of our
pseudo-volume limited catalogues are compared with the modelled
distribution in Fig.~\ref{fig:absmag_cat_zdist}, where good agreement
is demonstrated. Obviously, we can still calculate the average
absolute magnitude including K-corrections for each sample, and it is
this magnitude that we use to parametrise our bias model.

The distribution of absolute magnitude (without K-correction) against
redshift for the SDSS DR5 sample of galaxies is presented in
Fig.~\ref{fig:z_vs_R}. We plot ${\rm M}_r$ against redshift for 5\% of
the galaxies in the combined sample, randomly selected. The main
galaxy volume limited catalogues are delineated by the boxes in this
plot, and the upper and lower apparent magnitude limits of the main
galaxy survey are also shown.

For each pseudo-volume limited catalogue we have calculated the power
spectrum using the same method applied to our final catalogue in
Section~\ref{sec:pk_method}, with the exception that a uniform bias
model was applied -- we treat all galaxies as having $b=1$, so the
relative amplitudes of the recovered power are not affected by an
input bias model. Some of these power spectra are plotted in
Fig.~\ref{fig:pk_test2}. The window function for each sub-catalogue
was calculated as for our final catalogue, and we have fitted the
power spectrum amplitude over $0.01<k<0.2\hompc$ using
window-convolved models with approximately the correct large-scale
shape. The power spectra recovered from the different subsamples are
correlated, and we have not estimated their relative errors, as would
be strictly required when comparing their relative
amplitudes. Instead, we simply measure the average and standard
deviation of the difference between measured and model power spectra
over the range of scales of interest, and calculate the bias from
this. Consequently, the biases are not optimally determined, and their
errors do not include the effect of cosmic variance for the regions of
the catalogues that do not overlap, and will probably therefore
under-estimate the true error. However, as we show in test (5) of
Section~\ref{sec:tests}, we do not need to know the relative biases to
high precision as the resulting power spectrum is not sensitive to the
exact form of this correction.

\begin{figure}[tb]
  \plotone{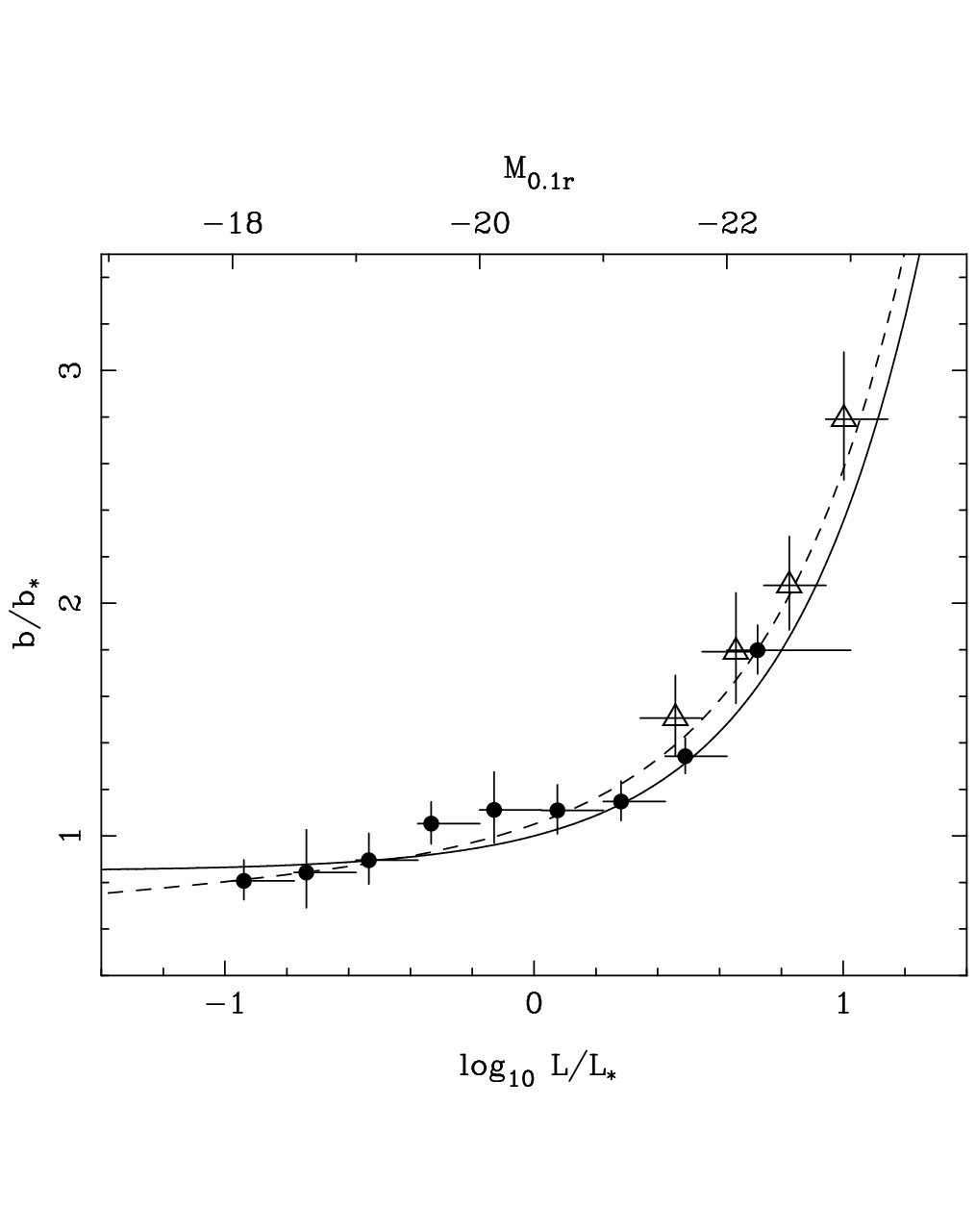}
  \caption{The bias of the SDSS main galaxies (solid circles with
    1-$\sigma$ errors) and LRGs (open triangles with 1-$\sigma$
    errors) as a function of ${\rm M}_{^{0.1}r}$. Horizontal errors
    show the range of luminosities analysed in each sample (ignoring
    K-corrections), while the data point is placed at the weighted
    average K-corrected luminosity. The relation suggested by
    \citet{norberg01}, $b/b_*=0.85+0.15L/L_*$, where ${\rm
      M}_*=-20.44$ \citep{blanton03b} is shown by the solid line. The
    dashed line shows the alternative formula of \citet{tegmark04},
    predicting a steeper increase in bias for luminous galaxies. As we
    are only interested in the relative normalisation of the bias, we
    allow $b_*$ to vary to fit the data for the different formulae,
    meaning that the curves do not cross at
    $L=L_*$. \label{fig:b_vs_R}}
\end{figure}

The relative biases measured from the pseudo-volume limited catalogues
are plotted (solid circles) in Fig.~\ref{fig:b_vs_R} as a function of
the average K-corrected absolute magnitude. The smooth shape matches
the overall shape of the simple formula of \citet{norberg01}
$b/b_*=0.85+0.15L/L_*$ (solid line), with ${\rm M}_*=-20.44$
\citep{blanton03b}. However, the addition of an extra term
$b/b_*=0.85+0.15L/L_*+0.04({\rm M}_*-{\rm M}_{^{0.1}r})$
\citep{tegmark04,zehavi05b} allows for a sharper increase in bias with
luminosity, which is a better fit to the data.

When measuring the power spectrum from the final combined galaxy
catalogue (see Section~\ref{sec:pk_method} for a description of the
method), we only need to know the averaged properties of the expected
bias at each spatial location. Consequently, even though we expect
galaxy bias to depend on colour as well as luminosity, a simple
luminosity-bias relation can still be used provided that the
catalogues from which the relation is derived contain the same
distribution of galaxy colours in each luminosity bin as galaxies of
that luminosity in the combined galaxy sample. For our pseudo-volume
limited subcatalogues of the main sample galaxies, we only exclude
galaxies that lie beyond the redshift limits applied for each
catalogue (See Fig.~\ref{fig:z_vs_R}). Because the catalogues are
relatively narrow in magnitude, $\Delta{\rm M}_r=0.5$, they contain
between 67\% and 77\% of the total number of galaxies within the
chosen magnitude limits (this number fluctuates because of the
changing steepness of the number counts shown in
Fig.~\ref{fig:z_vs_R}). Consequently, each sub-catalogue should have
approximately the same galaxy colour distribution as all galaxies of
that luminosity in the main sample. At high luminosities, the galaxies
are predominantly LRGs as shown in Fig.~\ref{fig:gmr_vs_r}, while at
low luminosities they are formed of a mix of red and blue galaxies.

\begin{figure}[tb]
  \plotone{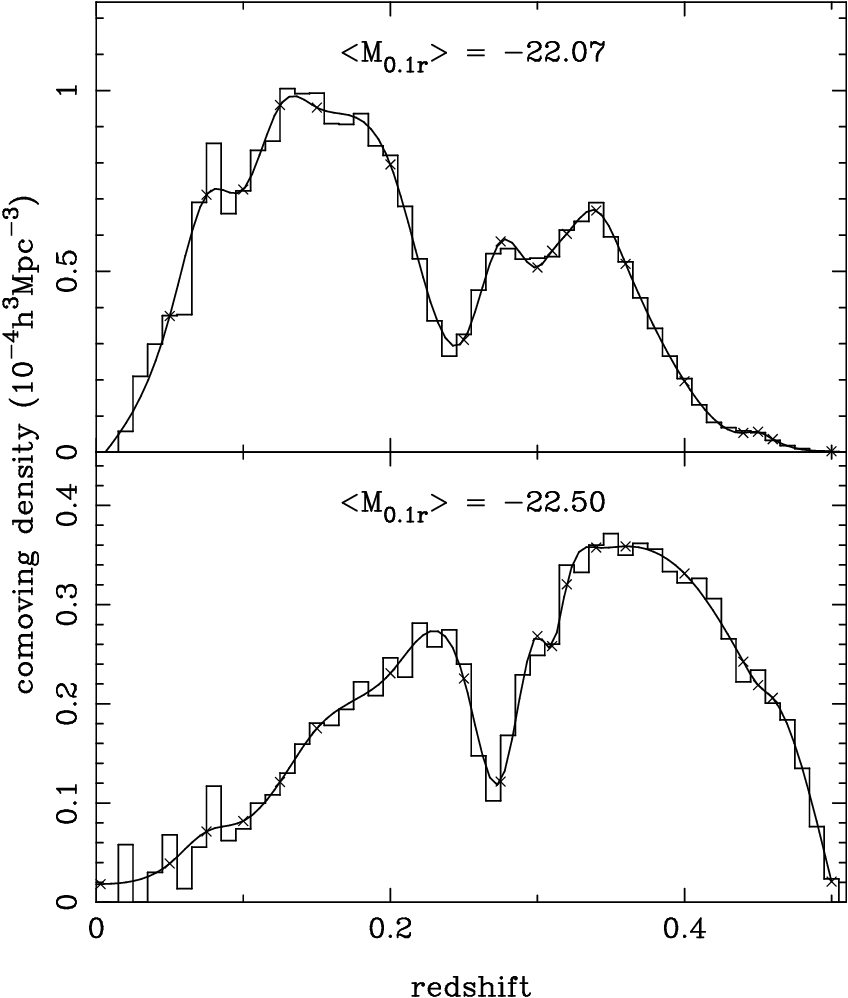}
  \caption{The number density as a function of redshift of two of our
    magnitude selected LRG catalogues (solid histograms). The
    distribution is fitted by a cubic spline (smooth curve) with the
    positions of the nodes (crosses) determined as described in the
    text. \label{fig:lrg_cat_zdist}}
\end{figure}

\subsection{LRGs}  \label{sec:bias_lrg}

The DR5 LRG sample is of sufficient size that the clustering can now
be measured as a function of luminosity as for the main galaxy
sample. The distribution of LRG magnitudes is plotted against redshift
in Fig.~\ref{fig:z_vs_R}, showing that the LRGs form a natural
extension of the main galaxy sample to higher redshifts. However,
given the more complicated selection function of the LRGs, modelling
the redshift distribution is not as straightforward as for the main
galaxy sample. Consequently, for sub-catalogues of LRGs selected as a
function of luminosity, we have fit the number density as a
function of redshift with the cubic spline fit as described in
Section~\ref{sec:sel_rad} for the total LRG sample. Example fits are
plotted for two of our LRG sub-catalogues in
Fig.~\ref{fig:lrg_cat_zdist}. Because of the increased volume covered
by the LRGs, although there are relatively few galaxies, the effect of
cosmological structure is small, and we do not expect to remove power
by fitting to the redshift distribution in this way.

The relative bias measurements for the LRGs were calculated from these
subsamples as for our main galaxy pseudo-volume limited catalogues,
and are plotted in Fig.~\ref{fig:b_vs_R} compared with the main galaxy
sample measurements. As can be seen, the bias increases with
luminosity for the LRGs as for the main galaxies, and the high
luminosity data match the simple formula of \citet{tegmark04}. We do
see a difference between LRG and main galaxy bias at low luminosities,
because the colour cuts applied to select these LRGs will remove blue
galaxies that are in the main galaxy sample (see the colour
distribution plotted in Fig.~\ref{fig:gmr_vs_r}), matching the
findings of \citet{zehavi05a}. This colour dependence is not included
in our bias model, so we simply exclude the lowest LRG luminosity bin
from our final combined sample, removing LRGs with
${\rm M}_{^{0.1}r}>-21.8$. We also exclude all high luminosity galaxies
${\rm M}_{^{0.1}r}<-23.0$ (whether main galaxy or LRG), where an expected
high bias would have a strong effect on the power spectrum.

We calculate the relative bias of galaxies using the K-corrected
absolute magnitudes, assuming a flat $\Lambda$ cosmology with
$\Omega_M=0.3$. The large-scale relative amplitudes of the power
spectra of our galaxy subsamples are only weakly dependent on the
model chosen to convert from redshift to comoving
distance. Consequently, even though we may assume a different
cosmology when analysing the full galaxy sample, we choose to still
use the bias as a function of K-corrected absolute magnitudes
calculated assuming a flat $\Lambda$ cosmology with $\Omega_M=0.3$. We
do not change the bias model as a function of cosmology.

\section{MEASURING THE POWER SPECTRUM}  \label{sec:pk_method}

\begin{figure}[tb]
  \plotone{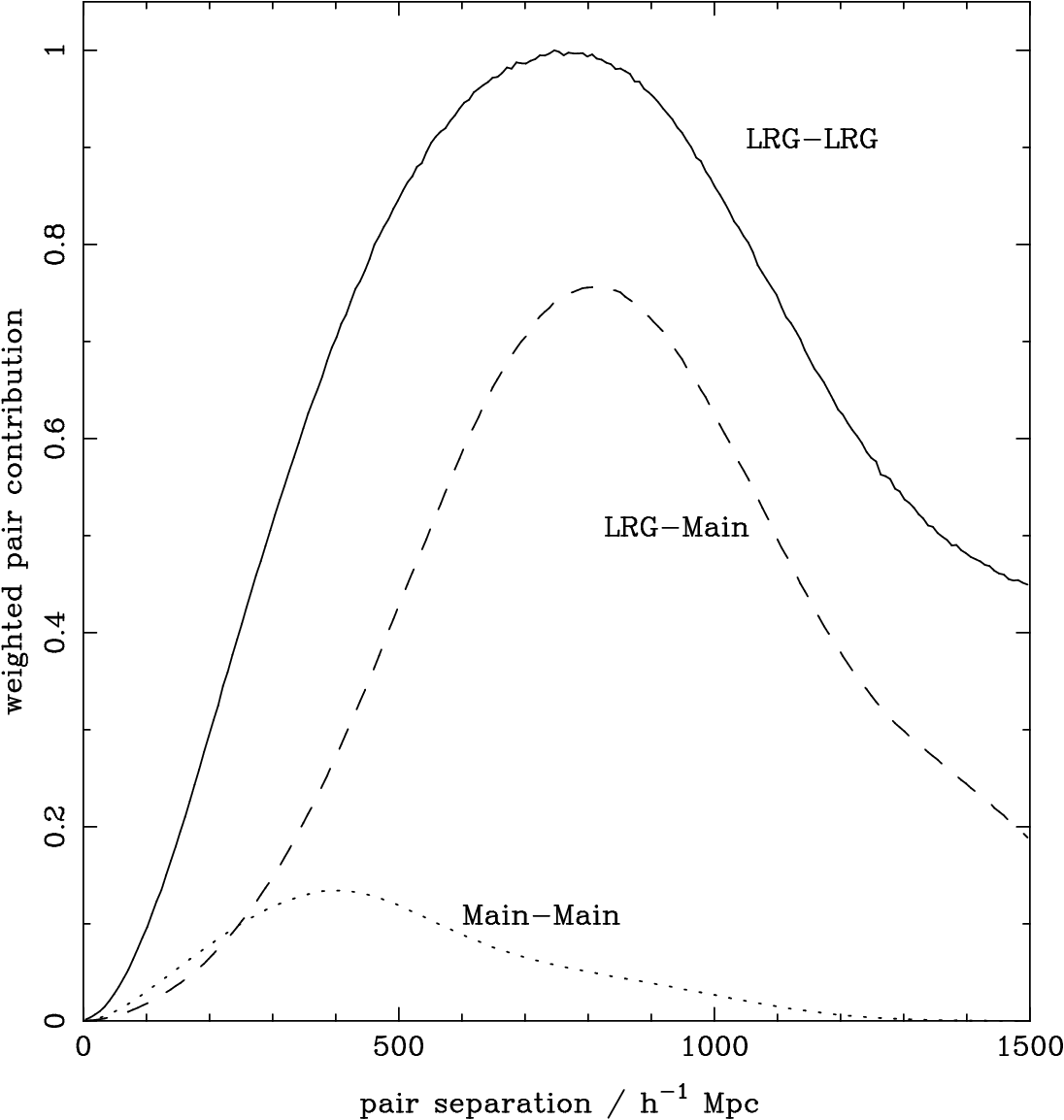}
  \caption{Weighted pair counts from LRG-LRG (solid line), main-main
    (dotted line), and LRG-main (dashed line) pairs. This plot shows
    how well the correlation function can be measured as a function of
    scale from these different samples, and is directly related to the
    power spectrum measurement at a particular scale
    (e.g. \citealt{tegmark06}). The LRG-LRG pairs dominate the
    measurement on all scales, although there is a significant
    contribution from pairs formed of a LRG and a main sample
    galaxy.\label{fig:lrg_main_compare}}
\end{figure}

\begin{figure}[tb]
  \plotone{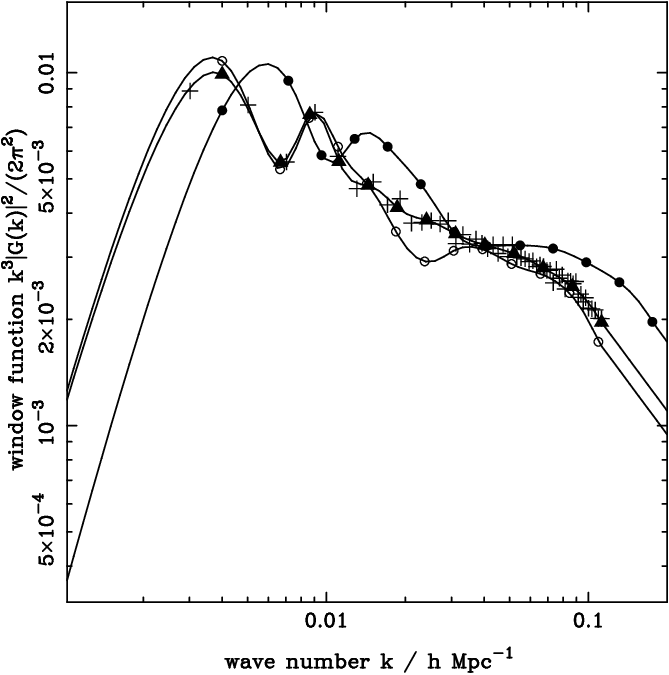}
  \caption{Cubic spline fits to the spherically averaged window
    functions in Fourier space (solid lines) with nodes given by open
    circles for the SDSS DR5 LRG subsample, solid circles for the main
    galaxy subsample and solid triangles for the combined LRG and main
    galaxy sample. For clarity, we only plot the measured window
    function power for the combined sample (crosses). In order to
    highlight small scale structure caused by the inhomogeneous nature
    of the survey, we multiply $|G(k)|^2$ by $k^3$. \label{fig:win_func}}
\end{figure}

The SDSS galaxy power spectrum has been calculated using the Fourier
based method of \citet{PVP}. Given a model for the average relative
bias at each location in the survey, this method extends the method of
\citet{FKP} to remove the effect of differential bias from the
recovered power spectrum. Additionally, this average bias is used to
optimally weight the galaxies -- more luminous galaxies are stronger
tracers of the underlying density field, contain more information
about the fluctuations, and therefore should be up-weighted compared
with less luminous galaxies.

The \citet{PVP} method was used by \citet{cole05} to analyse the
distribution of galaxies in the 2dFGRS, but there are some differences
between our analysis and that of \citet{cole05}, so we now provide a
brief description of the method applied to the SDSS highlighting these
differences. The primary change is that, because the sample of
galaxies extends to redshift $0.5$, we have dropped the common
assumption of using a single cosmological model in order to convert
redshifts to comoving radial distances. Instead, we recalculate the
power spectrum for different flat $\Lambda$ cosmological models. We
will sometimes wish to distinguish these cosmologies from those used
to create model power spectra. When we do this, we will refer to the
model used to calculate comoving distances as the ``cosmological
distance model''. Consequently, the data do not compress into a single
power spectrum, and there is no single power spectrum resulting from
our analysis.

At its heart, the Fourier method provides the simplest way to
calculate something approximating the galaxy power spectrum. The
galaxies are decomposed onto a grid, the grid is Fourier transformed,
and the amplitude of the Fourier modes measured. One key complication
is that weights are applied to each galaxy to optimally
reduce the error in the recovered power. Assuming a model for the
galaxy bias, the weights applied to a galaxy at location $\r$ with
expected bias $b'$ are those derived in \citet{PVP},
\be
  w(\r,b') = 
    \frac{(b')^2(\r)\bar{P}(k)}
         {1+\int db\,\langle n(\r,b)\rangle b^2\bar{P}(k)},
    \label{eq:w}
\ee
where $\bar{P}(k)$ is an estimate of the (unbiased) power spectrum,
and $\langle n(\r,b)\rangle$ is the expected density of galaxies as a
function of space and bias. In the analysis presented in this paper we
assume $\bar{P}(k)={\rm constant}=5000\,h^3Mpc^{-3}$, for
simplicity. This does not have a strong effect on the accuracy of the
recovered power spectrum. The relative weighted contributions from
pairs of LRGs, main galaxies and LRG-main galaxy pairs are plotted as
a function of scale in Fig.~\ref{fig:lrg_main_compare}. As can be
seen, although the LRGs dominate the analysis on all scales there is a
significant contribution to the weighted pair counts from main
galaxy-LRG pairs. Had we allowed $\bar{P}(k)$ to be reduced on small
scales (as would be optimal), then the higher density regions would
contribute more to the weighted pair counts and the main sample
galaxies would have dominated the distribution on these smaller
scales.

Given a weight $w_i$ and expected bias $b_i$ for each galaxy, the
overdensity field can be written
\be
  F(\r) = \frac{1}{N} \left[ 
    \sum_{\rm gal} \frac{w_i}{b_i}
    - \int db\, \frac{\langle w(\r,b)n(\r,b)\rangle}{b} \right],
  \label{eq:field}
\ee
where $N$ is a normalisation constant
\be
  N\equiv\left\{\int d^3r\left[\int db\,
    \langle w(\r,b)n(\r,b)\rangle \right]^2\right\}^{1/2},
  \label{eq:N}
\ee
and $\langle w(\r,b)n(\r,b)\rangle$ is the expected weighted
distribution of galaxies as a function of bias $b$ and location
$\r$. We choose to model this field using a random catalogue with
points selected at the mean galaxy density $\langle n(\r)\rangle$,
using the fits outlined in Section~\ref{sec:sel_rad}. We use a random
catalogue containing 10 times as many points as we have galaxies.
Because we only need to determine the integral of the average of the
weighted and bias corrected density, we do not need to assign a
luminosity to each point in the random catalogue and calculate a bias
from this. Instead, we simply calculate weights and biases for the
random catalogue by fitting the average radial values of $w_i$ and
$w_i/b_i$ in the galaxy catalogue as a function of redshift using
cubic splines. The weights in the random catalogue are renormalised,
compared with the weights applied to the galaxies so that $\int
F(\r)\,dr=0$, thereby matching the total weighted number density in
galaxy and random catalogues.

\begin{figure}[tb]
  \plotone{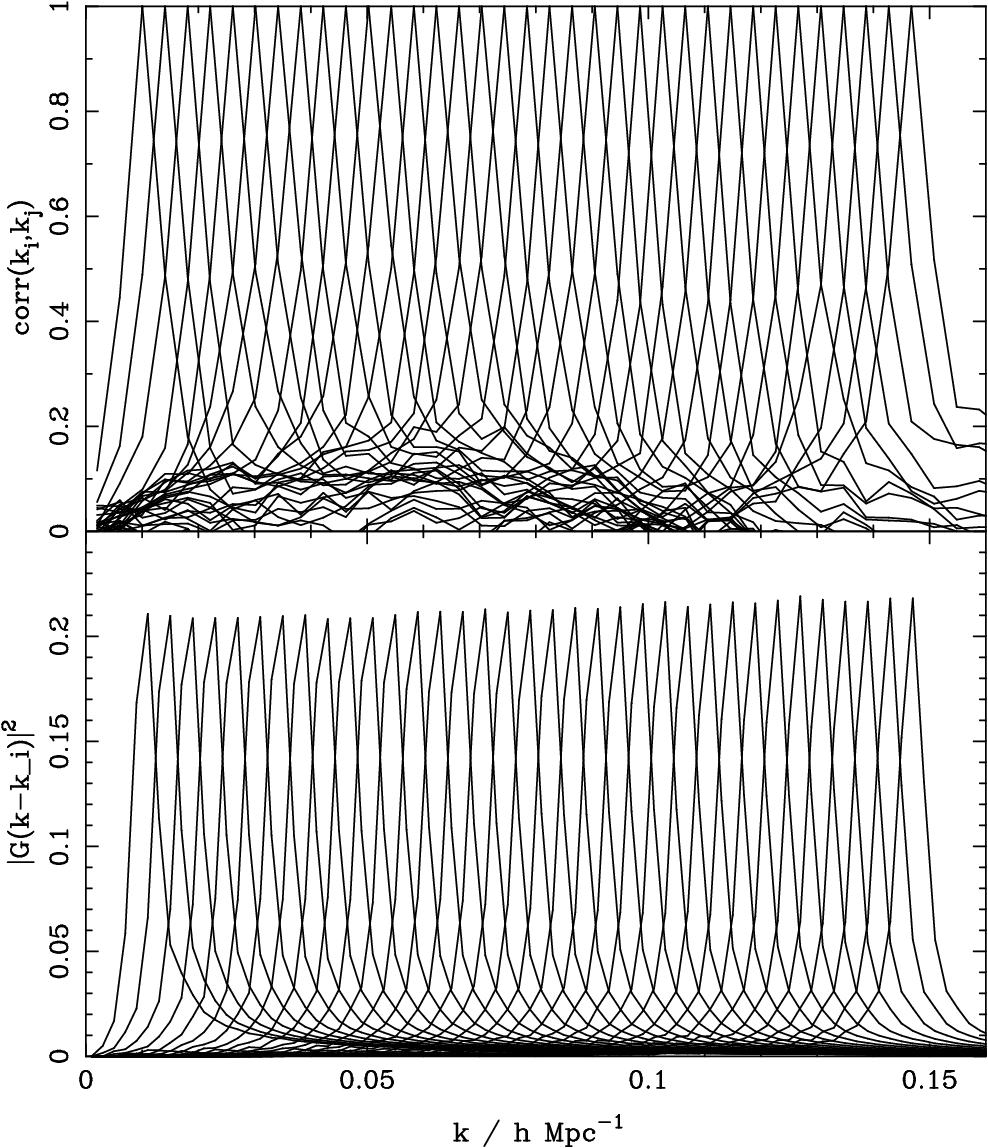}
  \caption{Top panel: correlations between data values calculated from
    2000 log-normal catalogues, assuming a flat $\Lambda$ cosmology
    with $\Omega_M=0.24$ for the cosmological distance model. Denoting
    the covariance between two power spectrum values $P(k_i)$ and
    $P(k_j)$ as ${\rm cov}(k_i,k_j)=\langle P(k_i)P(k_j)\rangle$, then
    we plot the correlation between the two measurements given by
    ${\rm cov}(k_i,k_j)/[{\rm cov}(k_i,k_i){\rm
      cov}(k_j,k_j)]^{1/2}$. For presentation, we have calculated the
    correlations plotted after matching the power spectra amplitudes
    recovered from the log-normal catalogues. This removes any
    normalisation error and only shows correlations induced by the
    window function. The covariance matrices used to calculate
    likelihoods were calculated from the raw power spectra calculated
    from the log-normal catalogues and therefore include the error in
    the overall normalisation. For $0.01<k<0.15\hompc$, we see that
    the correlation between data points is $<0.33$ for
    $|k_i-k_j|>0.01\hompc$. Bottom panel: The normalised window
    function for each of our binned power spectrum values with
    $0.01<k<0.15\hompc$. Each curve shows the relative contribution
    from the underlying power spectrum as a function of $k$ to the
    measured data values plotted in Fig.~\ref{fig:pk_bigplot}. The
    normalisation is such that the area under each curve is
    unity. \label{fig:kcorr}}
\end{figure}

\begin{figure}[tb]
  \plotone{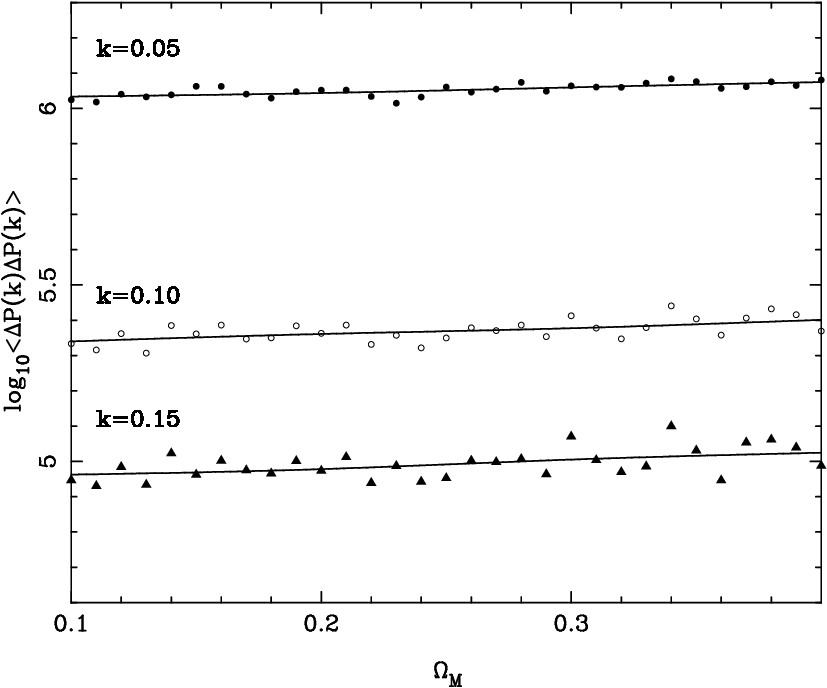}
  \caption{The amplitudes of three of the diagonal covariance matrix
    elements, estimated for flat $\Lambda$ cosmological models with
    different $\Omega_M$, from 2000 log-normal catalogues. These are
    plotted as a function of $\Omega_M$ with
    $\Delta\Omega_M=0.01$. The noise in the individual calculations is
    clear; this scatter has been minimised by fitting the data with a
    smooth cubic spline fit shown by the solid lines. We do not plot
    the error on the covariance matrix (the error in the error) for
    clarity. \label{fig:cov_mat_fit}}
\end{figure}

\begin{figure*}[tb]
  \plotone{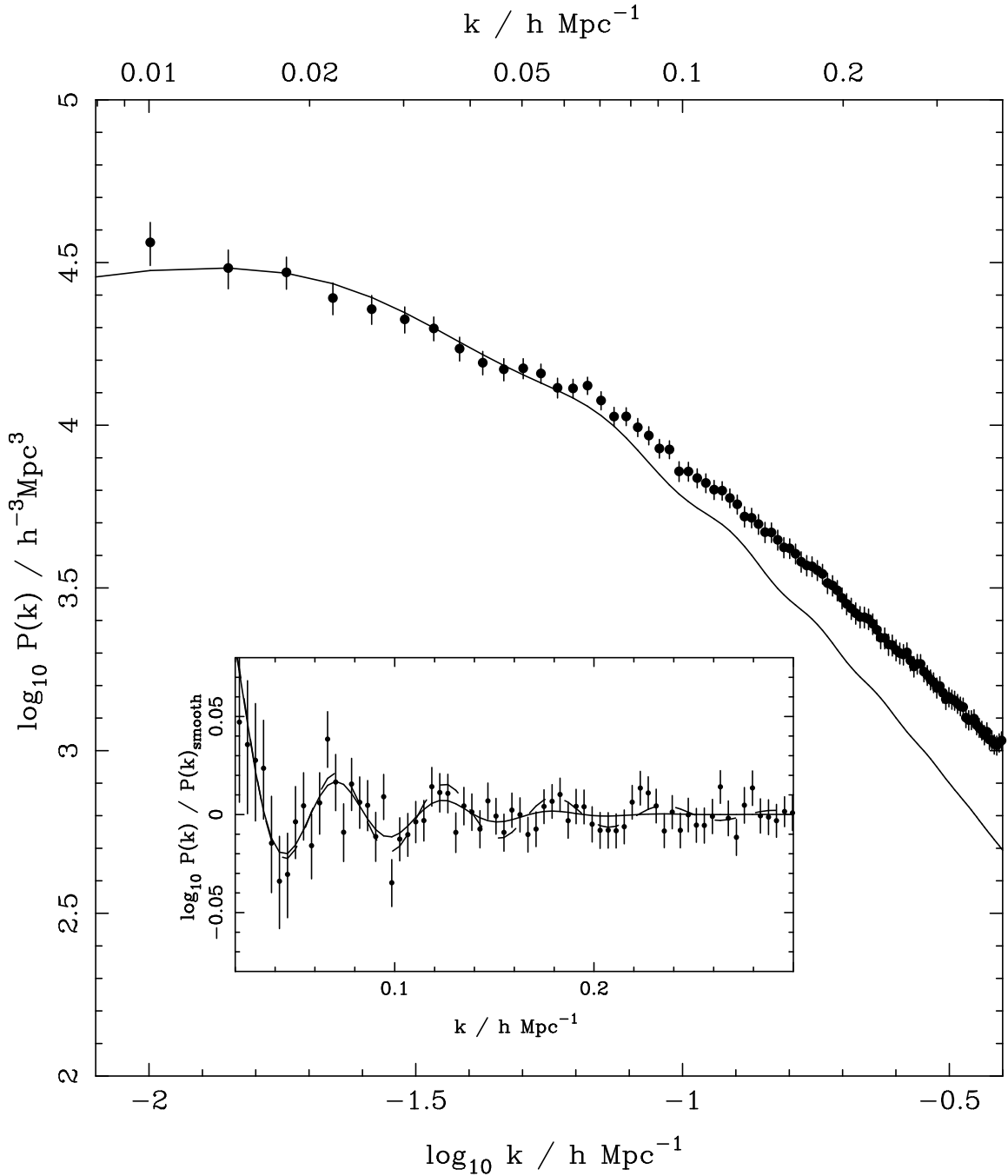}
  \caption{The redshift-space power spectrum recovered from the
    combined SDSS main galaxy and LRG sample, optimally weighted for
    both density changes and luminosity dependent bias (solid circles
    with 1-$\sigma$ errors). A flat $\Lambda$ cosmological distance
    model was assumed with $\Omega_M=0.24$. Error bars are derived
    from the diagonal elements of the covariance matrix calculated
    from 2000 log-normal catalogues created for this cosmological
    distance model, but with a power spectrum amplitude and shape
    matched to that observed (see text for details). The data are
    correlated, and the width of the correlations is presented in
    Fig.~\ref{fig:kcorr} (the correlation between data points drops to
    $<0.33$ for $\Delta k>0.01\hompc$). The correlations are smaller
    than the oscillatory features observed in the recovered power
    spectrum. For comparison we plot the model power spectrum (solid
    line) calculated using the fitting formulae of
    \citet{eisenstein98,eisenstein06}, for the best fit parameters
    calculated by fitting the WMAP 3-year temperature and polarisation
    data, $h=0.73$, $\Omega_M=0.24$, $n_s=0.96$ and
    $\Omega_b/\Omega_M=0.174$ \citep{spergel06}. The model power
    spectrum has been convolved with the appropriate window function
    to match the measured data, and the normalisation has been matched
    to that of the large-scale ($0.01<k<0.06\hompc$) data. The
    deviation from this low $\Omega_M$ linear power spectrum is
    clearly visible at $k\gs0.06\hompc$, and will be discussed further
    in Section~\ref{sec:omm_from_pk}. The solid circles with 1$\sigma$
    errors in the inset show the power spectrum ratioed to a smooth
    model (calculated using a cubic spline fit as described in
    \citealt{percival06b}) compared to the baryon oscillations in the
    (WMAP 3-year parameter) model (solid line), and shows good
    agreement. The calculation of the matter density from these
    oscillations will be considered in a separate paper
    \citep{percival06b}. The dashed line shows the same model without
    the correction for the damping effect of small-scale structure
    growth of \citet{eisenstein06}. It is worth noting that this model
    is not a fit to the data, but a prediction from the CMB
    experiment. \label{fig:pk_bigplot}}
\end{figure*}

The statistic that we use to compare models to the data is the shot
noise subtracted power spectrum of the overdensity field $F(\r)$,
measured in a series of $k$-space bins. The expectation value of this
statistic is
\be
  \langle |F(\k)|^2 - P_{\rm shot} \rangle
    = \int\frac{d^3k'}{(2\pi)^3}P(\k')|G(\k-\k')|^2,
 \label{eq:expFkFk}
\ee 
where we average $|F(\k)|^2$ over all $k$-space directions. The shot
noise is
\be 
  P_{\rm shot} = \sum_{\rm gal} \frac{w_i^2}{b_i^2} +
     \sum_{\rm ran} \left\langle \frac{w(\r,b)}{b}\right\rangle^2,
\ee 
and the window function $|G(\k)|^2$ is the normalised power in a
Fourier transform of 
\be 
  G(\r) = \int db\,\langle n(\r,b)w(\r,b)\rangle.  
\ee

The spherically averaged window functions from the LRGs, the main
galaxies and the combination of main galaxies and LRGs from the SDSS
DR5 sample are compared in Fig.~\ref{fig:win_func}. The large volume
probed by the LRGs means that the $k$-space window is small compared
with that calculated for the main galaxies. The window function from
the combined LRG and main galaxy sample is close to that of the LRGs,
although the main galaxies do provide more pairs of galaxies at
intermediate scales, smoothing the structure within the window
function. Spline fits to the window functions are used to numerically
determine the effect of the window on a model power spectrum. For a
smooth power spectrum, the features in the window function are
relatively unimportant compared with the overall shape. The
correlations induced by the window for the combined main galaxy and
LRG sample on the binned power spectrum are plotted in
the bottom panel of Fig.~\ref{fig:kcorr}.

The recovered power spectrum values are assumed to be distributed as a
multi-variate Gaussian, and we estimate the covariance matrix of this
Gaussian distribution using log-normal catalogues \citep{coles91}. For
each of 31 flat cosmological distance models with
$0.1\le\Omega_M\le0.4$ and $\Delta\Omega_M=0.01$, we have created 2000
log-normal catalogues (using the method described in
\citealt{cole05}). The distribution of galaxies in these catalogues
was calculated using the appropriate cosmological distance model,
while the power spectrum was calculated using a linear CDM model (see
Section~\ref{sec:model}) with parameters chosen to approximately match
the amplitude and shape of the recovered power for
$0.01<k<0.15\hompc$. As we will see in Section~\ref{sec:omm_from_pk},
this means that the model power is lower than the power calculated
from the data on scales $k<0.02\hompc$, so the errors calculated from
the log-normal catalogues are probably slightly overestimated on these
scales. The diagonal elements of the covariance matrix calculated in
this way are based on the effects of cosmic variance and shot
noise. The off-diagonal covariance matrix elements include both the
effect of the $P(k)$ window functions (bottom panel of
Fig.~\ref{fig:kcorr}) and the mode coupling induced by non-linear
evolution, to the extent that the latter is adequately described by
the log-normal approximation. The correlations induced by these
effects are shown in the top panel of Fig.~\ref{fig:kcorr}.

The noise in a single covariance matrix element would not normally be
noticed when calculating parameter constraints, as it would affect all
models in the same way. However, when we use different covariance
matrices to test different cosmological models these errors can become
important. Interestingly, if the power spectrum values have a Gaussian
distribution, then our estimates of the elements of the covariance
matrix will be drawn from a Wishart distribution, the same
distribution formed by temperature and polarisation CMB power spectra
in an all-sky survey \citep{percival06a}. We minimise the effect of
this noise by smoothing each element in the set of covariance matrices
using a separate 4-node cubic spline with nodes at
$\Omega_M=0.1,0.2,0.3,0.4$. Three examples of this fit are plotted in
Fig.~\ref{fig:cov_mat_fit}, for the diagonal elements of the
covariance matrix at $k=0.05,0.1,0.15\hompc$. There is a clear general
trend that the expected error in the power increases with the matter
density. This is caused by the changing comoving distance--redshift
relation, which means that a smaller volume is predicted for the
survey assuming $\Omega_M=0.4$, rather than $\Omega_M=0.1$.

The Fourier transforms used in this paper were performed on $512^3$
grids with a varying box size, where we only consider modes that lie
between $1/4$ and $1/2$ of the Nyquist frequency for each box. The
smoothing effect of the galaxy assignment is corrected as described in
\citet{cole05}. We have compared with both larger $1024^3$ and smaller
$256^3$ Fourier transforms, and find no evidence for systematics
induced by the practicalities of the Fourier transforms.

Assuming a flat cosmological distance model with $\Omega_M=0.24$,
matched to the parameters recovered from the 3-year WMAP CMB data, the
recovered SDSS power spectrum is plotted in
Fig.~\ref{fig:pk_bigplot}. Because of the bias-model correction, the
normalisation of this power spectrum is matched to that of $L_*$
galaxies, where ${\rm M}_*=-20.44$ \citep{blanton03b}. The precision
with which this power spectrum is measured, particularly on large
scales, is impressive. In Fig.~\ref{fig:pk_bigplot}, we also plot the
linear matter power spectrum predicted from the best-fit WMAP
parameters, normalised to match the data on scales
$0.01<k<0.06\hompc$. This shows good agreement in the shape of the
power spectrum on these scales, but deviates on scales
$k\gs0.06\hompc$. Significant non-linear corrections to the matter
power spectrum are only expected for $k\gs0.15\hompc$ \citep{smith03},
and large-scale redshift space distortions are commonly treated as
being scale-independent on these large scales. Consequently, it has
been common practice to assume that the shape of the power spectrum
recovered from galaxy surveys matches the linear matter power spectrum
shape on scales $k\ls0.15\hompc$
\citep{percival01,tegmark04}. However, the SDSS power spectrum has
greater power on small scales than the power spectrum predicted by the
3-year WMAP data.

\begin{figure}[tb]
  \plotone{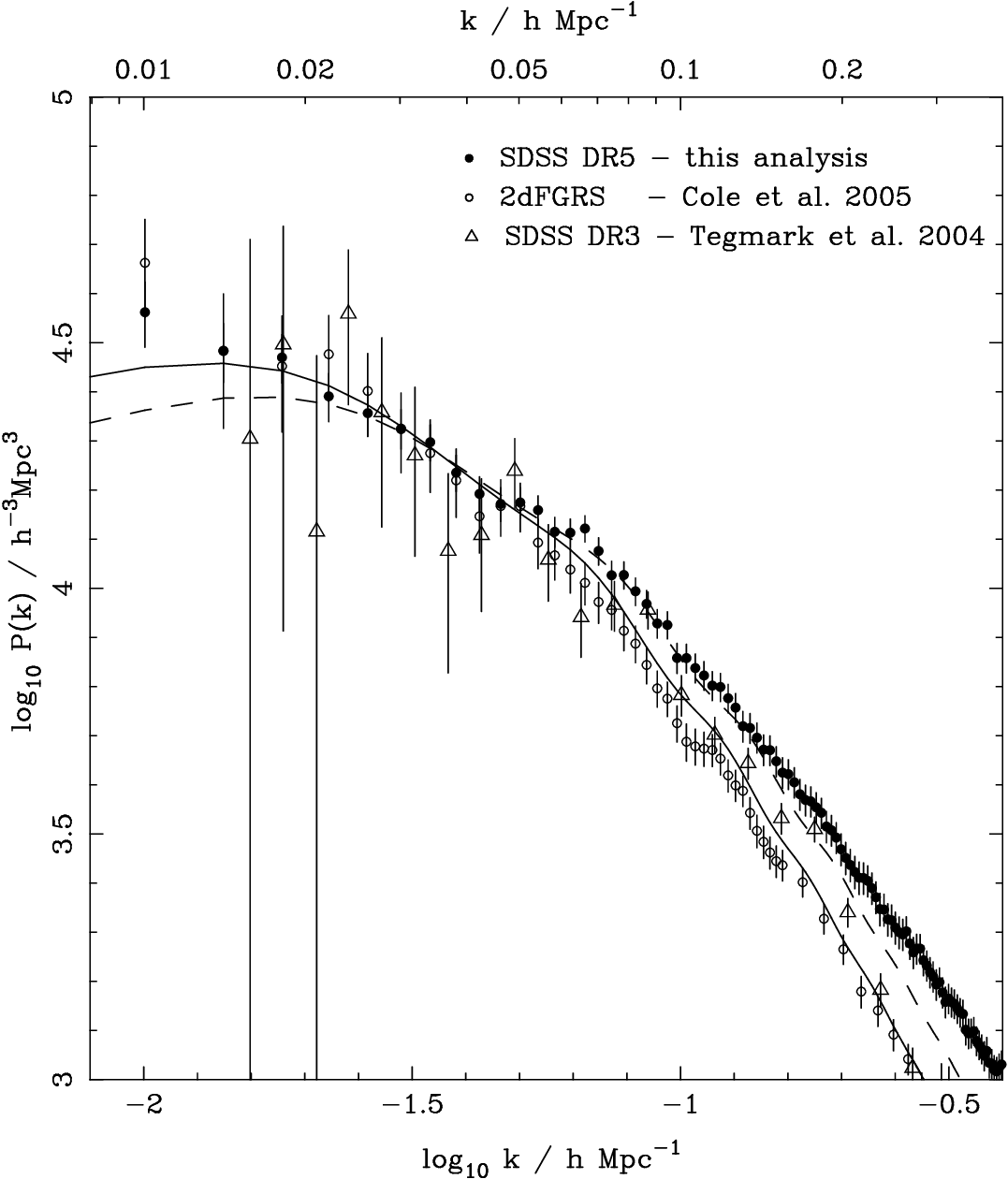}
  \caption{The recovered SDSS DR5 power spectrum plotted in
    Fig.~\ref{fig:pk_bigplot} compared with the previous SDSS
    real-space Main galaxy power spectrum calculated for the DR3
    sample (from Table~3 of\citealt{tegmark04}), and the 2dFGRS
    redshift-space galaxy power spectrum of \citet{cole05}. The data
    were corrected for the effects of the different window functions
    by calculating the multiplicative effect on a theoretical model
    with approximately the correct power spectrum shape. The
    normalisation of the previous data have been matched to that of
    the new power spectrum on large scales $k<0.06\hompc$. The shape
    of the power spectra agree well on these scales, but predict
    different power spectrum amplitudes on smaller scales. The solid
    and dashed lines show two linear CDM model power spectra, plotted
    as in Fig.~\ref{fig:pk_cmpr_mod}. \label{fig:pk_cmpr_previous}}
\end{figure}

In Fig.~\ref{fig:pk_cmpr_previous}, we compare the recovered power
spectrum with previous power spectra calculated from the SDSS DR3 main
galaxy sample \citep{tegmark04} and 2dFGRS \citep{cole05}.  The error
bars on the DR5 data points are much smaller than those on the DR3
data points in part because of the greater sky coverage of DR5, and in
(greater) part because of the inclusion of LRGs in the sample. It is
also worth noting that the procedure used by \citet{tegmark04}
constructs $P(k)$ estimates that are nearly independent, while our
estimates are correlated as shown in Fig.~\ref{fig:kcorr}.  After
corrections for the window functions and differences in the overall
normalisation, we see that the large-scale ($k\ls0.06\hompc$) shape of
the power spectrum recovered from the SDSS is well matched to that
recovered from the complete 2dFGRS. On smaller scales
($k\gs0.06\hompc$), our SDSS power spectrum of the combined main
galaxy and LRG sample reveals more power than that recovered from the
2dFGRS, while the previous power spectrum recovered from the SDSS DR3
main galaxy sample lies between the two. These differences will be
discussed further in Section~\ref{sec:discussion}. The power spectrum
analysis of the SDSS DR4 LRGs by \citet{huetsi06} also showed a
consistent power spectrum shape, although the current analysis
supersedes this in a number of ways. We analyse a larger sample of
galaxies and include an average bias model to correct for the fact
that we're analysing a mixed sample of galaxies. We have also
carefully constructed the angular mask of the survey, and have
provided a number of tests of the analysis method. Consequently we do
not directly compare with the power spectrum derived in this work.  In
the next section we analyse the shape of the recovered SDSS DR5 power
spectrum in detail, looking at the constraints it provides on the
matter density.

The 31 sets of power spectra, window functions, and covariance
matrices for the combined main-LRG sample, each computed using a
different $\Omega_M$ for cosmological distance calculations can be
obtained from \url{http://www.dsg.port.ac.uk/$\sim$percivalw/}, or upon
request from WJP. To use these data to compare the relative likelihood
of two model power spectra, one should choose for each model the data
set with the closest $\Omega_M$ value, convolve the model $P(k)$ with
the window functions provided, and calculate the likelihood for each
model from the tabulated $P(k_i)$ as a multi-variate Gaussian using
the corresponding covariance matrix.

\section{MODELLING THE GALAXY POWER SPECTRUM} \label{sec:model}

We calculate linear CDM model power spectra using the fitting formulae
of \citet{eisenstein98}, including a correction for the damping
of the baryon oscillations due to low-redshift small-scale structure
\citep{eisenstein06}. The damped power spectrum $P_{\rm damped}(k)$ is
assumed to be given by
\be
  P_{\rm damped}(k) = g P_{\rm lin}(k)+(1-g)P(k)_{\rm smooth},
\ee
where $P_{\rm lin}$ is the standard linear power spectrum, and
$P(k)_{\rm smooth}$ is the same power spectrum without the baryon
acoustic oscillations. The parameter $g=\exp(-k^2\sigma^2/2)$ is a
Gaussian damping term, and we assume $\sigma=10\mpcoh$ for the
spherically averaged redshift-space power spectrum that we measure
\citep{eisenstein06}.

There are three effects that distort the observed galaxy power
spectrum from the linear matter power spectrum. These are non-linear
structure growth, redshift-space distortions and galaxy bias. Note
that we take non-linear structure growth to correspond to the overall
behaviour of the matter in the universe, rather than the effect of the
collapse of small-scale structures on the galaxy power spectrum, which
may be different, depending on how galaxies trace the mass. In this
paper we will only be concerned with scales where the matter in the
universe is still expected to be predominantly in the linear regime
$k\ls0.15\hompc$ \citep{smith03}, so the effect of non-linear
structure growth is small. For the best-fit 3-year WMAP parameters
\citep{spergel06}, the \citet{smith03} fitting formulae predict an
increase in power at $k=0.15\hompc$ due to non-linear effects of
8\%. Increasing the value of $\sigma_8$ from $0.77$ to $0.9$ predicts
a 10\% increase due to non-linear effects at $k=0.15\hompc$. In
contrast, if the linear power spectrum predicted by the WMAP
experiment is normalised to the data on scales $0.01<k<0.06\hompc$,
then the measured SDSS power spectrum is $40\%$ greater than this
model at $k=0.15\hompc$.

While redshift space distortions only depend on the distribution of
mass -- the galaxies effectively act as test particles within the
gravitational potential, the strength of galaxy bias is predicted to
depend on galaxy properties. In particular, there is some theoretical
work that suggests that we might expect a scale-dependent galaxy bias,
even on scales $k\ls0.15\hompc$. \citet{seljak01} suggest that within
the halo model \citep{seljak00,peacock00,cooray02}, there are two
effects that may cause a scale-dependent bias for pairs of galaxies in
different halos, which is particularly strong for LRGs (where the
galaxies only occupy the most massive halos): if there is an
additional Poisson selection of halos to be populated, then there may
be a Poisson term in the resulting power spectrum in addition to the
standard shot noise term due to the finite number of galaxies (that is
subtracted in the \citealt{FKP} method). Such a Poisson term would
show up on large scales where the power spectrum has a lower
amplitude.  Additionally, for the most massive halos, the bias is a
strong function of the halo mass. Small changes in the average mass of
the halos occupied by the galaxies as a function of the scale probed
could lead to a scale dependent galaxy bias.

Using halo model calculations, \citet{cole05} introduced a simple two
parameter model (hereafter the $Q$-model) to correct for the effects
of galaxy bias and redshift space distortions, with
\be
  P_{\rm gal} = \frac{1+Qk^2}{1+Ak} P_{\rm lin},
\ee
where $A=1.4$ and $Q=4.0$ was suggested for the redshift-space 2dFGRS
power spectrum by fitting halo model simulations. For the real-space
power spectra, halo model simulations instead predicted that $A=1.7$
and $Q=9.6$. On large scales, the different values of $A$ mean that
the redshift-space distortions increase the power spectrum slope (so
the ratio of small-scale to large-scale power is higher). On
small-scales the converse is true, and the redshift-space distortion
decrease in the amplitude of the small-scale power. Between
$0.01<k<0.15\hompc$, if the model power spectra are normalised on
large scales, then this model predicts a 7\% lower redshift-space
power spectrum at $k=0.15\hompc$ compared with the real-space power
spectrum. This is very similar to the offset between linear and
non-linear power discussed above using the \citet{smith03} fitting
formulae. We see that the real-space power spectrum for the 2dF
galaxies predicted by the halo model tracks the non-linear increase in
power, while the redshift-space distortions effectively cancel the
non-linear increase in power leading to a redshift-space galaxy power
spectrum shape that more closely matches that of the linear matter
power spectrum.

\begin{figure}[tb]
  \plotone{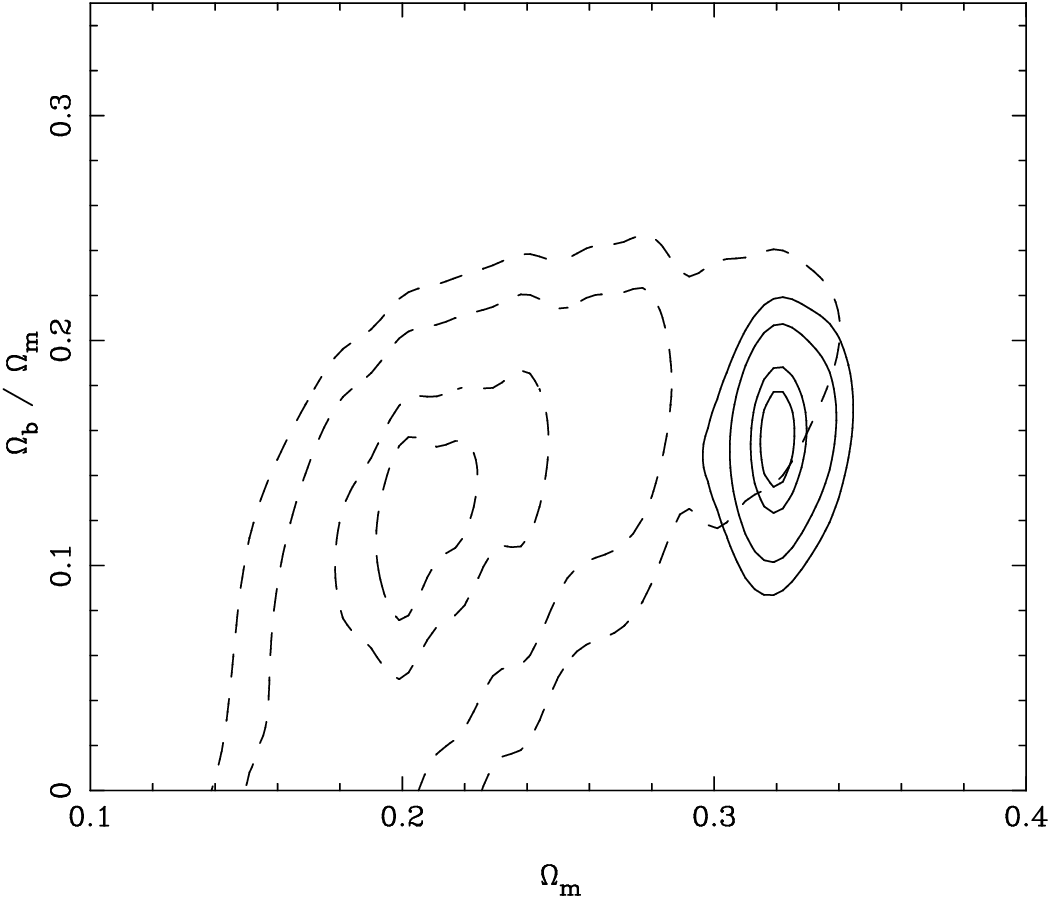}
  \caption{Likelihood contours calculated from fitting the power
    spectrum shape with CDM models. $n_s=0.96$, and $h=0.73$ are
    assumed and we vary $\Omega_M$ and the baryon fraction. For each
    value of $\Omega_M$ tested, we recalculate the power spectrum, the
    window function and the expected error on the power so the
    cosmological distance model matches the power spectrum
    model. Dashed contours were fitted for $0.01<k<0.06\hompc$, and
    solid contours for $0.01<k<0.15\hompc$. Contours are plotted for
    $-2\ln{\cal L}=1.0,\,2.3,\,6.0,\,9.2$, corresponding to
    one-parameter confidence of 68\%, and two-parameter confidence of
    68\%, 95\% and 99\% for a Gaussian
    distribution. \label{fig:like.allOm}}
\end{figure}

\begin{figure}[tb]
  \plotone{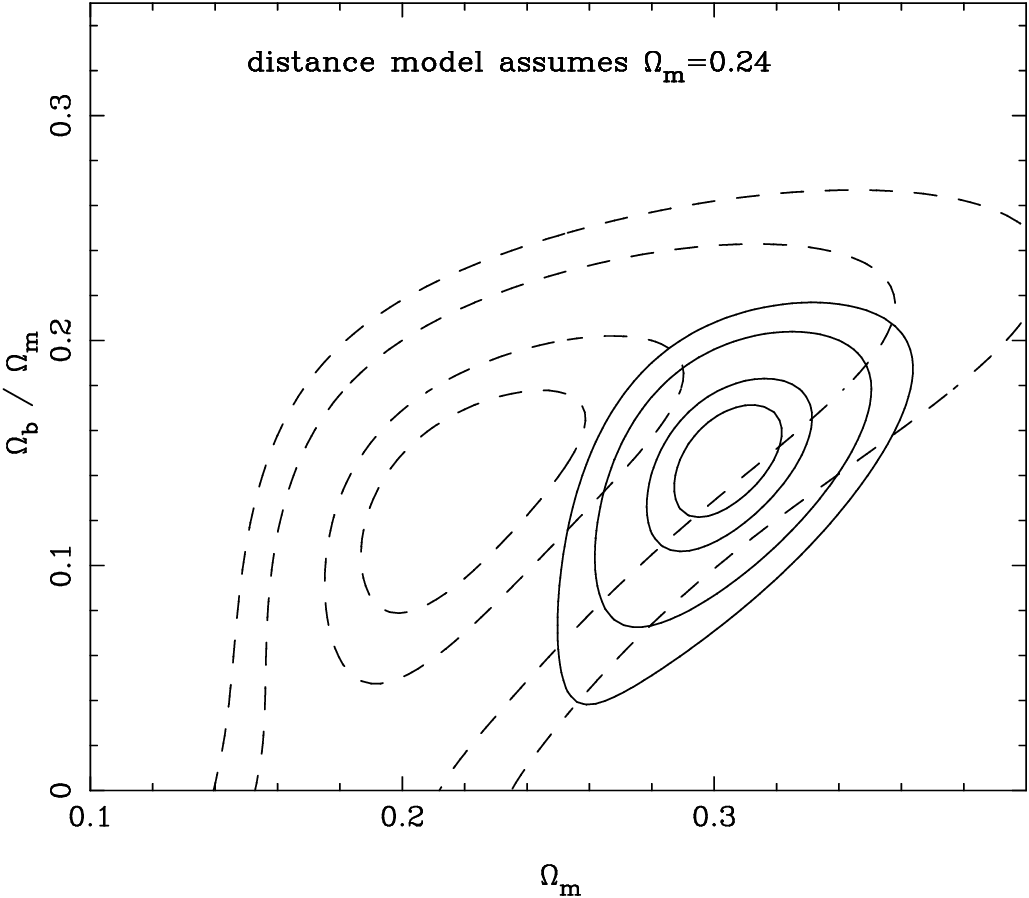}
  \caption{As Fig.~\ref{fig:like.allOm}, but now only considering a
    single power spectrum, calculated assuming a flat $\Omega_M=0.24$
    cosmology to determine comoving galaxy distances from the measured
    galaxy redshifts. \label{fig:like.Om0.24}}
\end{figure}

The two parameter $Q$-model is used by \citet{tegmark06} in their
analysis of the SDSS LRG power spectrum; here $A=1.7$ was fixed
(corrections for redshift-space effects are included in the method),
but $Q$ was allowed to vary to fit the data, with a best-fit value of
$Q\simeq26$ determined from the power spectrum fits in this work. The
different values of $Q$ for the SDSS LRGs and 2dFGRS galaxies reflect
a change in shape in the power spectrum, with the LRGs being more
biased on small scales relative to large scales. Obviously, in
addition to this change in shape, the SDSS LRGs are more biased at any
scale than the 2dFGRS galaxies. With the values of $A$ and $Q$
appropriate for the LRGs, the $Q$-model correction to the linear power
spectrum varies between $0.98$ and $1.01$ over $0.01<k<0.06\hompc$,
and increases on smaller scales to $1.31$ at $k=0.15\hompc$. It is
worth noting that, while the Q-model was not designed for such highly
biased populations, the modelling presented in \citet{tegmark06}
suggests that it might still fit the clustering of these luminous
galaxies.

Rather than assume such a prescription when fitting models to the data
in this paper, we take a step back and instead consider the
observational evidence that such a correction is required. First, we
test the hypothesis that the observed large-scale power spectrum shape
matches that of a CDM linear matter power spectrum over the scales
where the matter is predicted to be approximately in the linear
regime. We do this by assuming that the observed galaxy power spectrum
does match a linear CDM model, and then consider if this results in a
contradiction in the cosmological parameters derived fitting to
different scales. This hypothesis effectively assumes that the
contributions to the power spectrum from non-linear structure growth
and redshift-space distortions cancel, while galaxy bias is
scale-independent. Any contradiction would show (from a single data
set), that this cannot be the case. Second, the parameter values
$Q\simeq26$ for the SDSS LRGs and $Q\simeq4$ for the 2dFGRS galaxies
in the $Q$-model suggest that we should expect a strong change in the
shape of the galaxy power spectrum as a function of galaxy
properties. We split the large SDSS DR5 sample into subsamples to look
for such effects on large scales $0.01<k<0.15\hompc$ in
Sections~\ref{sec:tests} \&~\ref{sec:pk_lum}.

\begin{figure}[tb]
  \plotone{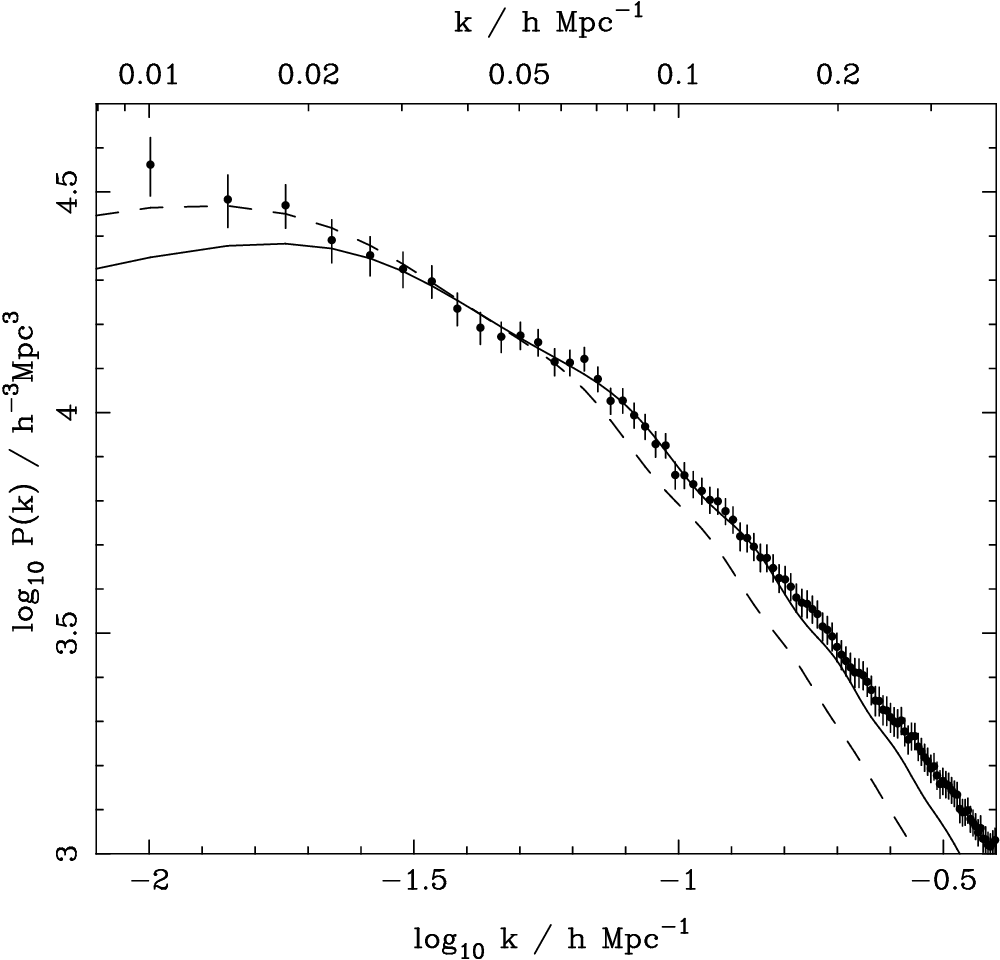}
  \caption{The power spectra recovered from the combined SDSS DR5
    sample assuming $\Omega_M=0.24$ to convert from redshifts to
    comoving distances (solid circles with 1-$\sigma$ errors) compared
    with the best fit CDM models. The dashed line shows the best-fit
    model over $0.01<k<0.06\hompc$, while the solid line shows the
    best-fit model over $0.01<k<0.15\hompc$. The two fits are
    discrepant at approximately
    2-3\,$\sigma$. \label{fig:pk_cmpr_mod}}
\end{figure}

\section{$\Omega_M$ FROM THE POWER SPECTRUM SHAPE} 
\label{sec:omm_from_pk}

In this section we fit CDM models to the SDSS DR5 power spectrum data
assuming that the observed power spectrum matches the shape of a
linear CDM model.  We have calculated a grid of model power spectra as
described in Section~\ref{sec:model}, allowing the matter density
$\Omega_M$ and baryon fraction $\Omega_b/\Omega_M$ to vary, assuming a
scalar spectral index of $n_s=0.96$, and a Hubble parameter
$h=0.73$. Varying $n_s$ causes a small change in the recovered value
of the matter density: following \citet{cole05}, the effect can be
approximated by $(\Omega_M h)_{\rm true} = (\Omega_M h)_{\rm apparent}
+ 0.3(1-n_s)$. Similarly, the effect of a non-zero neutrino fraction
would change the recovered value by approximately $(\Omega_M h)_{\rm
  true} = (\Omega_M h)_{\rm apparent} + 1.2(\Omega_\nu/\Omega_M)$. For
each value of the matter density tested, we have recalculated the
window function from the geometry of the sample using the appropriate
comoving distance-redshift relation. This window function is used to
convolve the model power, and is used to correct for the loss of power
due to the normalisation of the overdensity field: because the total
expected number of galaxies is unknown, the normalisation of the
random catalogue was matched to the galaxy catalogue so that $\int
F(\r)\,dr=0$ (see Eq.~\ref{eq:field}, and the subsequent
discussion). The effect of this on the data is to subtract a multiple
of the window function, so that $P_{\rm data}(0)=0$. We therefore
subtract a multiple of the window function from the model power
spectra (after convolution with the appropriate window) so that
$P_{\rm model}(0)=0$. This is the equivalent of the ``integral
constraint'' correction required for measured correlation
functions. The likelihoods of the model power spectra are then
calculated assuming that the data form a multi-variate Gaussian
distribution with the appropriate covariance matrix, calculated as
described in Section~\ref{sec:pk_method}.

The resulting likelihood surfaces are plotted in
Fig.~\ref{fig:like.allOm}.  Contours are plotted for fits to two
different $k$-ranges, $0.01<k<0.06\hompc$ (dashed contours), and
$0.01<k<0.15\hompc$ (solid contours). As can be seen, the choice of
scales fitted makes a strong difference to the recovered best fit
parameters. On scales $0.01<k<0.06\hompc$, the shape matches that of a
low matter density cosmology with $\Omega_M=0.22\pm0.04$. Extending
the fit to smaller scales, the data prefer a higher matter density
$\Omega_M=0.32\pm0.01$. The fits are discrepant at the $2-3\sigma$
level. The corresponding marginalised baryon fractions are
$\Omega_b/\Omega_M=0.13\pm0.05$ and
$\Omega_b/\Omega_M=0.16\pm0.03$. On scales $0.01<k<0.06\hompc$, the
small scale damping does not strongly affect the recovered power and
we recover the same parameter constraints if we do not make the
small-scale damping correction to the model baryon acoustic
oscillations. Fitting to $0.01<k<0.15\hompc$ without the damping term
reduces the baryon fraction constraint slightly to
$\Omega_b/\Omega_M=0.15\pm0.02$.

The results from fitting the power spectrum shape are only weakly
dependent on the model assumed to calculate comoving distances from
galaxy redshifts. We have considered fitting general CDM models to the
single power spectrum calculated assuming a fixed cosmological
distance model with $\Omega_M=0.24$ (as plotted in
Fig.~\ref{fig:pk_bigplot}), and likelihood contours are presented in
Fig.~\ref{fig:like.Om0.24}, as for Fig.~\ref{fig:like.allOm}. The
general form of the contours remains consistent, although the
locations of the likelihood maxima do change -- for the fit to
$0.01<k<0.15\hompc$ we recover $\Omega_M=0.31\pm0.02$.

The best fit models are compared with the data in
Fig.~\ref{fig:pk_cmpr_mod}, which clearly shows why the discrepancy
arises. There is too much power on scales $0.01<k<0.06\hompc$ for the
$\Omega_M=0.32$ model at approximately 1\,$\sigma$. Conversely, for
the fit to the large-scale data, there is too much power on smaller
scales $k>0.06\hompc$.

\begin{figure*}[tb]
  \plotone{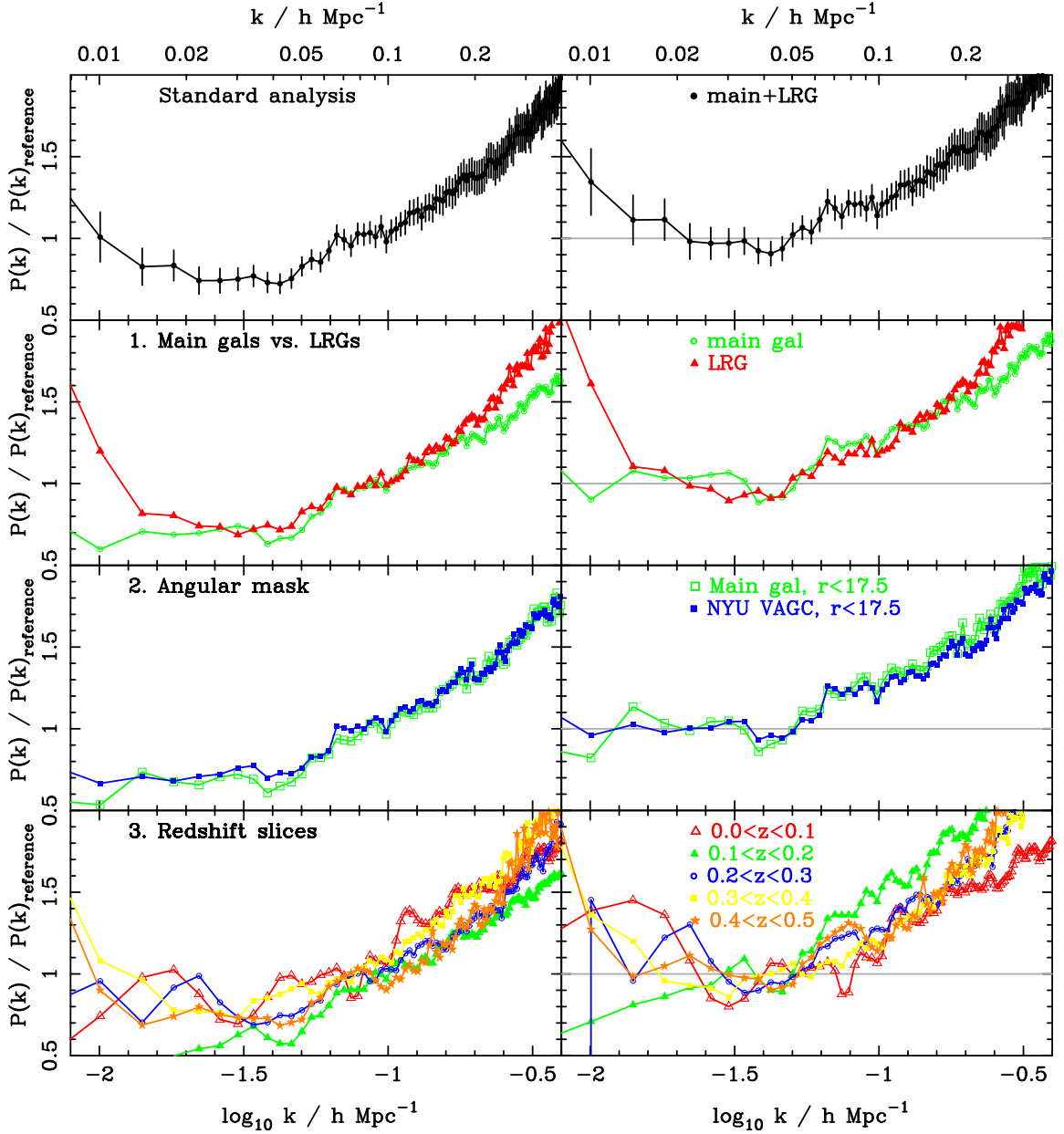}
  \caption{Power spectra calculated from the SDSS DR5 sample divided
    by a model linear power spectrum with $h=0.73$, $\Omega_M=0.18$,
    $n_s=0.96$ and no baryons (in order not to introduce extra
    oscillations in the plotted power spectra). In the left column, we
    simply divide by the raw model power spectrum, with a fixed
    normalisation matched to the final power spectrum from the
    combined LRG + Main galaxy sample.  In the right hand column we
    convolve the model power spectrum with the window function
    appropriate to the test being performed and correct for the loss
    of power due to the normalisation of the overdensity field, before
    calculating the ratio of the power spectra. The amplitude is
    allowed to vary to match the data on scales $0.01<k<0.06\hompc$,
    so we are only comparing the shapes of the power spectra recovered
    from the different tests in the right column. The top row shows
    the ratios calculated for our final power spectrum derived from
    the combined main galaxy and LRG catalogue. The other rows show
    power spectra calculated either for different subsamples of this
    catalogue, or using slightly different techniques, and are
    described in Section~\ref{sec:tests}. \label{fig:pk_test1}}
\end{figure*}
\begin{figure*}[tb]
  \plotone{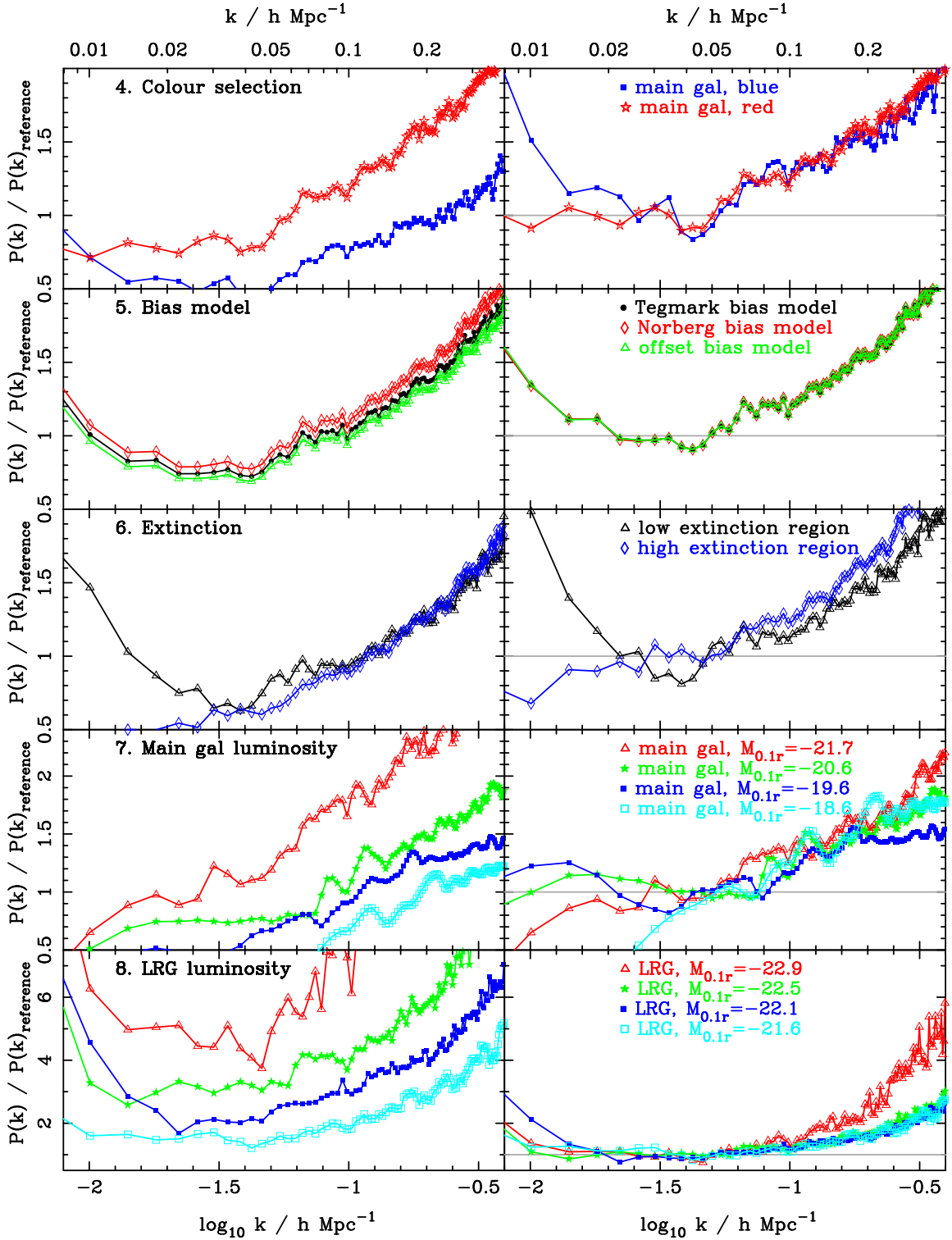}
  \caption{Continuation of Fig.~\ref{fig:pk_test1}. To accommodate the
    range of power spectrum normalisations in tests (7) and (8), the
    scale of the y-axis was changed in these
    panels. \label{fig:pk_test2}}
\end{figure*}

\section{POWER SPECTRA OF SUB-CATALOGUES AND ROBUSTNESS TESTS}  
  \label{sec:tests}

In this section we present a number of test power spectra, calculated
for variations of our method or for different subsamples of the
data. The power spectra were calculated using a flat cosmological
model with $\Omega_M=0.24$ to convert from redshifts to comoving
distance, and are compared in Figs.~\ref{fig:pk_test1}
\&~\ref{fig:pk_test2}. In the left column they are divided by a linear
power spectrum calculated with no baryons and with $\Omega_M=0.18$,
which has approximately the correct shape but no baryonic features
and, in the right column, to the same model convolved with the
appropriate window function for each power spectrum, and with matched
normalisation. For comparison we plot the power spectrum recovered
using our standard analysis method in the top row of
Fig.~\ref{fig:pk_test1}.

\begin{enumerate}

\item We have split the catalogue into (slightly overlapping) main
  galaxy and LRG subsamples, and have calculated power spectra for
  these samples using our standard method and our standard
  luminosity-dependent bias model. Fig.~\ref{fig:pk_test1} shows that
  the large-scale normalisation of the recovered power spectra match
  well compared with the relative average bias of the two
  populations. This shows that the bias model used is renormalising
  both power spectra to match that of an $L_*$ galaxy. There is weak
  evidence for a small normalisation offset, matching the small offset
  in the measured LRG bias compared with the model assumed (the open
  circles are slightly above the dashed line in
  Fig.~\ref{fig:b_vs_R}). We will test the bias model later, and show
  that small deviations in the bias assumed for the LRGs compared with
  the main galaxies do not affect the power spectrum calculated from
  the combined sample (test 5). On small scales the LRG power has a
  higher amplitude compared with the main galaxy power.

\item We test our angular mask by comparing recovered power spectra
  from samples of main galaxies selected to a limiting magnitude of
  $r<17.5$ using the angular and radial selection methods described in
  Section~\ref{sec:dr5}, with a similar catalogue obtained from the
  SDSS New York University Value Added Catalogue (NYU-VAGC;
  \citealt{blanton04a}). When analysing the NYU-VAGC sample we used
  their angular mask, which therefore matches the mask derivation
  previously used for SDSS team analyses
  \citep{tegmark04,eisenstein05,tegmark06}, although this calculation
  has obviously been updated to cover the SDSS DR5 region. There are
  small changes between the NYU-VAGC mask and our mask, due to
  slightly different selection criteria and regions covered, but in
  the areas of overlap they match well. We select a ``safe'' sample of
  main galaxies with a magnitude limit of $r<17.5$ in order to provide
  a sample with a single radial distribution, and avoid complications
  from differing faint survey limits. For the NYU-VAGC catalogue we
  use the model provided for the radial distribution of galaxies,
  which was calculated from fits to the luminosity function as
  described in \citet{tegmark04}. For our sample we calculate the
  radial distribution of galaxies as described in
  Section~\ref{sec:sel_rad}. Even though these independent methods
  vary substantially in design for both the angular and radial
  selection, as can be seen in Fig.~\ref{fig:pk_test1}, there are only
  very minor differences between the recovered power spectra.

\item We test the radial distribution assumed for the galaxy
  population by comparing the power spectra measured in 5 redshift
  slices of width $\Delta z=0.1$ through the combined sample of main
  galaxies and LRGs. The standard luminosity-dependent bias model is
  assumed, and appears to adequately renormalise the power spectra
  calculated from the different samples -- there is no evidence for a
  significant change in normalisation of the power as a function of
  redshift. However, the magnitude limited catalogues from which the
  bias model was derived have redshift increasing with luminosity. The
  derivation of the model is therefore coupled with any redshift
  evolution. Consequently, it is perhaps not surprising that the
  amplitudes of the power spectra recovered in different redshift
  slices are so similar, although it is still gratifying to see that
  this is correct. A simple model with constant galaxy clustering as a
  function of redshift and with redshift-independent
  luminosity-dependent bias is therefore sufficient to model the
  currently observed clustering, but is not necessarily a unique
  solution.

\item The main galaxy sample is bimodal in colour; we have decomposed
  this dataset into red and blue galaxy subsamples using a simple
  colour cut ${\rm M}_{^{0.1}g}-{\rm M}_{^{0.1}r}=0.8$ (see
  Fig.~\ref{fig:gmr_vs_r}). These subsamples were analysed in exactly
  the same way as our final combined sample. In particular, the
  redshift distribution fitting function of Eq.~\ref{eq:zfit} was
  found to still provide an adequate match to the observed redshift
  distribution. The power spectra for these subsamples were corrected
  using our bias model, so the change in normalisation of the power
  spectra of the red and blue subsamples demonstrates the additional
  colour dependent bias term that is not included in our average bias
  model. As can be seen, there is no obvious change in the shape of
  the power spectra as a function of colour.

\item We also consider a power spectrum calculated using the bias
  model of \citet{norberg01} rather than the bias model of
  \citet{tegmark04}. The primary effect is a change of normalisation,
  equivalent to a change in the definition of $b_*$. Once this is
  corrected in the right hand panel of Fig.~\ref{fig:pk_test2}, we see
  no significant change in the recovered power spectra. We also
  consider an ``offset'' bias model, where we assume a model bias for
  the LRGs given by $b/b_*=0.85+0.15L/L_*+0.08({\rm M}_*-{\rm
    M}_{^{0.1}r})$, but do not change the bias model for the main
  galaxies. If this model is used to calculate the power in the LRG
  and main galaxy catalogues separately, the recovered large-scale
  power spectrum amplitude recovered from the LRGs is reduced, and the
  LRG and Main galaxy power spectra are in better agreement (see test
  1). For the combined sample, (or the LRG or main galaxy samples if
  analysed independently), this change does not affect the overall
  shape of the power spectrum, giving us confidence that any error
  introduced by joining the two galaxy catalogues is not significant.

\item The angular coverage of the SDSS sample is now sufficiently
  large that we can split the catalogue as a function of Galactic
  extinction. The median $r$-band extinction in the sample is $0.065$,
  and we split into galaxies with a higher extinction correction, and
  galaxies with a lower extinction correction. The recovered power
  spectra are not independent, and we have not calculated relative
  errors for these data. However, the power spectra diverge on scales
  $k<0.02\hompc$, and are also slightly discrepant on scales
  $k=0.05\hompc$, which could indicate a systematic problem with the
  extinction corrections. A recent test of the number density of SDSS
  galaxies as a function of the \citet{schlegel98} Galactic extinction
  correction was performed by \citet{yahata06}. They find that the
  number density of galaxies increases with increasing extinction for
  SFD extinction values below 0.1 magnitudes in the $r$-band. The
  contamination of the far infrared brightness of the sky by
  background galaxies postulated as an explanation of this effect in
  this work, might also explain the observed difference between these
  power spectra.

\item The final tests presented in Fig.~\ref{fig:pk_test2} compare
  power spectra recovered for galaxies of different luminosity. For
  the main galaxies, we plot 4 of the 9 pseudo-volume limited
  catalogues described in Section~\ref{sec:bias}. Each catalogue is
  0.5 magnitudes wide (not K-corrected), and no bias model has been
  included. Consequently, the increase in the normalisation of the
  large-scale clustering as a function of luminosity is clear in the
  left column.

\item As for (7), but now comparing 4 LRG sub-catalogues, each of
  width 0.5 magnitudes (not K-corrected). Looking at the left column,
  the increase in the overall clustering strength with luminosity is
  clear. In the right column, we see that any change in the shape of
  the power spectrum as a function of luminosity is at a level
  significantly below the change in the large-scale normalisation. The
  possibility that the power spectrum does change shape on scales
  $0.01<k<0.15\hompc$ as a function of luminosity will be considered
  further in the next Section.

\end{enumerate}

\begin{figure}[tb]
  \plotone{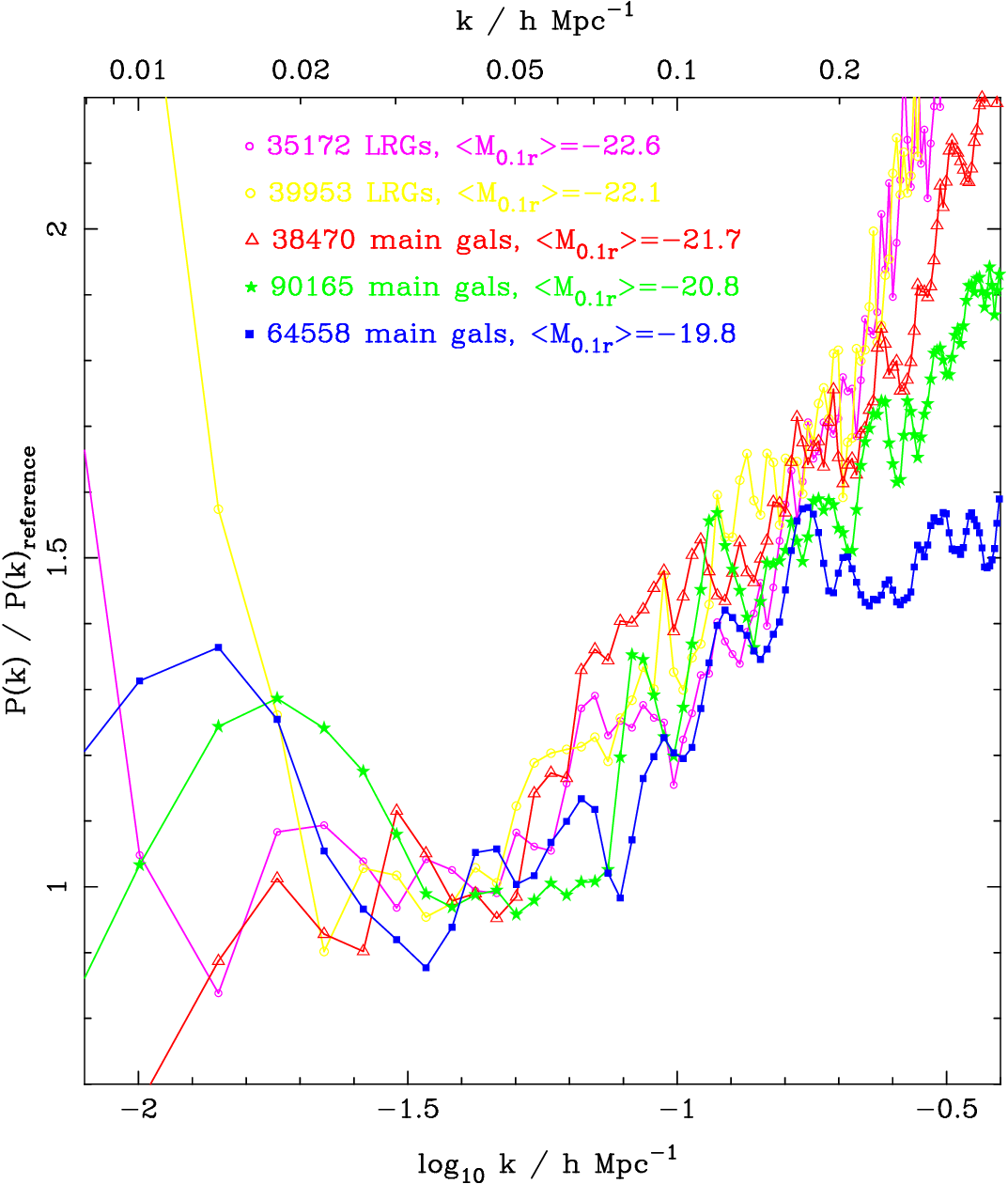}
  \caption{Power spectra calculated from LRG and main galaxy
    subcatalogues divided by a smooth CDM model matched to the
    subcatalogue (as in the right panels in
    Fig.~\ref{fig:pk_test1}). The amplitude of the CDM model was
    matched to the data over the narrow $k$-range $0.03<k<0.05\hompc$.
    For the main galaxies, 3 pseudo-volume limited catalogues of width
    1-magnitude covering $-22.5<{\rm M}_r<-19.5$ were created and
    analysed (as described in Section~\ref{sec:bias_maingal}). For the
    LRGs, the catalogue was split at approximately the median
    luminosity of the sample ${\rm
      M}_{^{0.1}r}=-22.3$. \label{fig:pk_vs_lum_big}}
\end{figure}

\section{OBSERVED CLUSTERING AS A FUNCTION OF LUMINOSITY} 
  \label{sec:pk_lum}

In this section, we expand on tests (7) and (8) presented in
Section~\ref{sec:tests} in order to consider whether there is any
evidence that the power spectrum changes shape as a function of
luminosity on scales $0.01<k<0.15\hompc$. In order to test this, we
have created three pseudo-volume limited subcatalogues of the main
galaxy sample, using the method described in
Section~\ref{sec:bias_maingal}, but now each of width 1-magnitude,
covering $-22.5<{\rm M}_r<-19.5$ (without K-correction). We have also
split the LRG sample into two catalogues at approximately the median
magnitude ${\rm M}_{^{0.1}r}=-22.3$. Power spectra have been
calculated for these five subcatalogues, and are plotted in
Fig.~\ref{fig:pk_vs_lum_big}, divided by a fiducial power spectrum
model matched to each power spectrum as in the right column of
Figs.~\ref{fig:pk_test1} \&~\ref{fig:pk_test2}.

With the amplitude of the data matched on large scales, on small
scales $k>0.2\hompc$, there is a clear hierarchy with more luminous
galaxies showing stronger small-scale clustering. In order to quantify
the effect on cosmological parameter estimation in the regime where
the matter is still linear ($k\ls0.15\hompc$), the important question
is ``on what scale do the power spectra start to deviate from each
other?''. Unfortunately, the data are insufficient to answer this
question in a quantitative way. For the samples split from the main
galaxy catalogue, there is perhaps a slight trend with the least
luminous galaxies having less power on scales $0.06<k<0.15\hompc$, but
this conclusion is fairly speculative.

\section{DISCUSSION} \label{sec:discussion}

We have analysed the clustering of galaxies in the largest sample (in
terms of number and volume covered) to date. We have shown that the
SDSS main galaxy and LRG samples can be naturally combined, because
both sets of galaxies, after careful luminosity selection, can be
described using the same average large-scale luminosity-dependent bias
model. Analysing the combination of data allows us to include
cross-correlation between the two data sets as well as internal
auto-correlations within either the LRG or main galaxy samples.

Because of the speed and simplicity of the Fourier method used, we
have been able to consider a number of tests of the recovered power
spectrum, and have eliminated a number of possible systematics from
our analysis. The only unusual result was a possible problem with the
Galactic extinctions; this requires further detailed analysis and
modelling that is outside the scope of our paper. In addition to
considering a larger data set than previously analysed, our analysis
extends previous work by considering the cosmological model to be
tested from the start of the analysis -- using this model to convert
from redshift to comoving distance, as well as comparing with the
resulting power spectrum.

In a companion paper we have considered the cosmological constraints
from the baryon acoustic oscillations observed in the power spectrum,
where we found $\Omega_M=0.250^{+0.042}_{-0.021}$ for flat $\Lambda$
cosmological models \citep{percival06b} when combined with the
observed baryon acoustic oscillation scale in the CMB
\citep{spergel06}. This paper has instead focused on trying to model
the overall shape of the power spectrum, paying particular attention
to systematic effects on the resulting matter density
constraints. Because of the accuracy with which we can now measure the
power spectrum on scales $k<0.05\hompc$ as a consequence of the large
volume covered by the LRG sample, we can test the hypothesis that the
shape of the redshift-space galaxy power spectrum matches the matter
on scales where the matter clustering is well modelled by linear
evolution from the seed perturbations ($0.01<k<0.15\hompc$). We find a
discrepancy between fits to $k<0.06\hompc$, and $0.06<k<0.15\hompc$,
albeit at only moderate significance. The data on scales
$0.01<k<0.06\hompc$ favour a model with $\Omega_M=0.22\pm0.04$, while
extending the $k$-range to $0.01<k<0.15\hompc$ increases the estimated
value to $\Omega_M=0.32\pm0.01$. Statistically, this discrepancy is
only at the 2-3$\sigma$ level.  It is worth noting that the excess
power on scales $k < 0.06\hompc$ relative to the $\Omega_M=0.32$ model
matches that recently found from analyses of the SDSS
photometric-redshift sample \citep{padmanabhan06,blake06}. This
suggests that the discrepancy is therefore not limited to our
analysis. The upper $k$ limit for the large-scale fit, $k_{\rm
  lim}=0.06\hompc$, was conservatively chosen based on the observed
shape of the power spectrum (Fig.~\ref{fig:pk_bigplot}). The noise in
the data means that there is nothing special in this choice, and we
could have split the scales fitted at $0.05<k_{\rm lim}<0.1\hompc$,
and obtained similar results.

Interestingly, the change in the recovered matter density with the
scales fitted matches the bimodality of previous matter density
constraints, with the best-fit model on scales $0.01<k<0.06\hompc$
matching that predicted by the WMAP 3-year data \citep{spergel06}, the
2dFGRS \citep{cole05}, and the results from the baryon acoustic
oscillations observed in the SDSS DR5 sample. In contrast, the fits to
$0.01<k<0.15\hompc$ match the results from early studies of the SDSS
data by \citet{pope04} and \citet{tegmark04}, suggesting that we are
recovering approximately the same overall shape on these scales as
these early analyses. The new data obviously constrain $\Omega_Mh$
with greater accuracy so, on the scales previously considered, the
significance of the offset between 2dFGRS and SDSS analyses has
increased. Additionally, the higher value of $\Omega_M=0.32\pm0.01$ is
now more discrepant with the (mean) WMAP constraint
$\Omega_M=0.234\pm0.034$ (from Table~2 of \citealt{spergel06}), and is
discrepant with the lower matter density recovered from the positions
of the baryon acoustic oscillations in the power spectrum
\citep{percival06b}.

The hypothesis being tested by these model fits is that the
contributions to the power spectrum from non-linear structure growth
and redshift-space distortions cancel, galaxy bias is
scale-independent, and that the matter clusters as expected in a
simple CDM model with an unbroken power spectrum. The results provide
a contradiction to this hypothesis with a significance of
2-3$\sigma$. We will now consider each of the elements in turn to try
to understand which assumption is breaking down. First, it is worth
noting that the effect of non-linear structure growth was discussed in
Section~\ref{sec:model} and shown to be too small to cause the
observed power spectrum distortion.

Small-scale redshift-space distortions act in the wrong direction for
the observed effect, and it would require a significant break-down of
the scale-independent increase in power spectrum amplitude predicted
by \citet{kaiser87} for linear infall observed at large distances to
give a scale-independent increase in the power spectrum amplitude and
cause the observed change in shape. N-body simulations, and the halo
model calculations performed for the 2dFGRS also found that the effect
of redshift space distortions on the shape of the power spectrum is
small on the scales of interest \citep{percival01,cole05}. The
consistency between the higher matter density favoured fitting to
$0.01<k<0.15\hompc$ in our study, previous studies of the real-space
power spectrum \citep{tegmark04}, and the power spectrum calculated
from sets of photometric-redshift LRGs also provide evidence against
redshift-space distortions producing the observed effects.

An unwelcome possibility is a systematic problem with one of the data
sets, or the analysis method, although the only discrepancy revealed
in the tests performed in Section~\ref{sec:tests} was between the
power spectra recovered using different Galactic extinction correction
cuts. However, this potential systematic could also explain
differences between 2dFGRS and SDSS power spectra if it predominantly
affects the SDSS galaxies; analysing the galaxy clustering in the low
extinction regions in the SDSS, for which \citet{yahata06} show that
the number density of galaxies does not behave as expected, produces a
slight excess of power on scales $k\simeq0.06\hompc$ (see
Fig.~\ref{fig:pk_test2}). Additionally, it is perhaps also worth
mentioning the heretical possibility that there is a problem with the
assumption that the matter clusters as expected in a CDM model with an
unbroken post-inflation power law spectrum. However, such a problem
would affect the 2dFGRS and SDSS galaxies in the same way, so a
further explanation would be required for this difference.

Perhaps the most simple explanation for this inconsistency is that the
luminous red galaxies that dominate our combined sample do not trace
the linear matter power spectrum as simply as other galaxies, and
there is some theoretical work that supports this assertion.  In order
to try to obtain evidence for scale-dependent galaxy bias, we have
analysed the shape of the SDSS power spectrum, particularly looking
for evidence of an increase in small-scale clustering power that
depends on galaxy properties. The luminosity-bias correction applied
as part of the Fourier method will only correct for the large-scale,
scale-independent bias affecting different luminosity galaxies. It is
not designed to correct for scale-dependent bias.  We have analysed a
number of subsets of our final catalogue in order to look for
scale-dependent bias, and find no significant change in shape if we
split our sample using the bimodal galaxy colour distribution, or if
we split in redshift; we recover similar power spectrum shapes,
analysing the data in 5 redshift slices out to $z=0.5$. There is weak
evidence for a change in the shape of the power spectrum for
$0.01<k<0.15\hompc$ when splitting the galaxies by luminosity, which
varies with the average $r$-band luminosity of the galaxy sample
analysed. However, the evidence for this is not conclusive. It is
worth noting that we clearly see the effects of colour and luminosity
on smaller scales $k>0.2\hompc$, in line with the results of
\citet{cole05}. If scale-dependent bias also affects large scales
then, in principle, this is testable by measurement of the bispectrum
on the same data set \citep{scoccimarro06}: the bispectrum shape and
scale dependence respond to bias in a way that differs from, and is
therefore not degenerate with, the power spectrum.

A change in power spectrum shape on scales $k<0.15\hompc$, as a
function of luminosity would provide a consistent picture when we
compare our recovered power spectrum with previous work. On scales
$k\ls0.06\hompc$, Fig.~\ref{fig:pk_cmpr_previous} shows that the shape
is consistent with the 2dFGRS power spectrum \citep{cole05}. If the
amplitudes of the power spectra are matched on these large scales,
then on smaller scales there is a progression from the low clustering
amplitude of the 2dFGRS galaxies, through the main galaxies of the DR3
SDSS sample analysed by \citet{tegmark04} to the higher clustering
strength of the combined main galaxy and LRG sample of this
analysis. There is no reason to suggest that this is not a natural
progression following the trend observed within the SDSS from galaxies
with a low $r$-band luminosity to those with a high $r$-band
luminosity. In this interpretation, the ``excess power'' on the
largest scales probed by the \citet{padmanabhan06} and \citet{blake06}
analyses of photometric LRGs is a consequence of comparing to an
incorrect reference power spectrum with high $\Omega_M$.  The true
value of $\Omega_M$ is lower, and the excess power is on small scales,
as in Fig.~\ref{fig:pk_bigplot} here.  The observed clustering was
calculated using a weighted average over all galaxy
pairs. Consequently, if we analyse the DR5 LRGs and main galaxies
separately and match the large-scale clustering amplitudes, we would
expect a progression from a lower small-scale amplitude of the main
galaxies, through the combined sample, to a higher clustering
amplitude of the LRGs. Fig.~\ref{fig:pk_vs_lum_big} shows that this is
indeed what we find.

The combination of scale-dependent bias, redshift-space distortions
and non-linear structure growth for a set of galaxies has previously
been matched by a simple fitting formulae (the $Q$-model) applied
after calculation of the power spectrum \citep{cole05}. However, it
remains to be seen if this simple prescription can adequately describe
galaxy bias for LRGs or for a mixed sample of galaxies with different
clustering properties; these issues will be considered in
\citet{tegmark06}. The analysis presented in our paper suggests that
this correction could become increasingly important for samples as the
average $r$-band galaxy luminosity increases. It is clear that, for
LRGs, the relation between galaxies and dark matter will need to be
carefully modelled in order to extract the maximum possible
information from the shape of the observed power spectrum.

\section{SUMMARY}

In summary, using the SDSS, the redshift-space galaxy power spectrum
is now known with sufficient accuracy on large scales to test the link
between galaxies and the underlying matter distribution within the
class of CDM models. If we assume a scale-independent bias between
galaxies and the mass, we find that no linear CDM model can fit the
data over $0.01<k<0.15\hompc$, with fits to $0.01<k<0.06\hompc$
suggesting a matter density that is $2-3\sigma$ from that derived
fitting to $0.01<k<0.15\hompc$. Perhaps the simplest explanation is
that the large-scale distribution of luminous red galaxies is affected
by scale-dependent bias, although we cannot rule out the alternative
possibility that this is due to a systematic effect. If we calculate
power spectra for subsets of galaxies selected using the $r$-band
luminosity, then we only see weak evidence for a change in shape of
the power spectrum on the appropriate scales, with galaxies that are
less luminous having a scale-dependent bias that is weaker and affects
smaller scales than more luminous galaxies. Such a bias model could
also explain the bimodality in matter densities calculated from other
data sets and from previous SDSS power spectra. The blue-selected
2dFGRS galaxies would be less affected by this scale-dependent bias
than the red-selected galaxies and bias modelling would be less
important when calculating cosmological constraints. Hence the simple
assumption of \citet{percival01} that the 2dFGRS galaxy power spectrum
has the same shape for $k<0.15\hompc$ as the matter power spectrum is
comparatively harmless. As the $r$-band luminosity of the sample
increases, the non-linear relation between the galaxies and dark
matter needs to be carefully modelled when providing cosmological
constraints from galaxy clustering even on relatively large scales
$k<0.15\hompc$

\acknowledgments

We thank the referee, Joss Bland-Hawthorn for helpful comments and
suggestions. WJP is grateful for support from a PPARC fellowship, and
RCN for a EU Marie Curie Excellence Chair.

Funding for the SDSS and SDSS-II has been provided by the Alfred
P. Sloan Foundation, the Participating Institutions, the National
Science Foundation, the U.S. Department of Energy, the National
Aeronautics and Space Administration, the Japanese Monbukagakusho, the
Max Planck Society, and the Higher Education Funding Council for
England. The SDSS Web Site is \url{http://www.sdss.org/}.

The SDSS is managed by the Astrophysical Research Consortium for the
Participating Institutions. The Participating Institutions are the
American Museum of Natural History, Astrophysical Institute Potsdam,
University of Basel, Cambridge University, Case Western Reserve
University, University of Chicago, Drexel University, Fermilab, the
Institute for Advanced Study, the Japan Participation Group, Johns
Hopkins University, the Joint Institute for Nuclear Astrophysics, the
Kavli Institute for Particle Astrophysics and Cosmology, the Korean
Scientist Group, the Chinese Academy of Sciences (LAMOST), Los Alamos
National Laboratory, the Max-Planck-Institute for Astronomy (MPIA),
the Max-Planck-Institute for Astrophysics (MPA), New Mexico State
University, Ohio State University, University of Pittsburgh,
University of Portsmouth, Princeton University, the United States
Naval Observatory, and the University of Washington.

Simulated catalogues were calculated and analysed using the COSMOS
Altix 3700 supercomputer, a UK-CCC facility supported by HEFCE and
PPARC in cooperation with CGI/Intel.

\setlength{\bibhang}{2.0em}

\label{lastpage}


\begin{thebibliography}{999}

  \bibitem[\protect\citeauthoryear{Abazajian et al.}{2003}]{abazajian03}
    Abazajian K., et al., 2003, AJ, 126, 2081

  \bibitem[\protect\citeauthoryear{Abazajian et al.}{2004}]{abazajian04}
    Abazajian K., et al., 2004, AJ, 128, 502

  \bibitem[\protect\citeauthoryear{Adelman-McCarthy et al.}{2006a}]{adelman06a}
    Adelman-McCarthy J., et al., 2006a, ApJS, 162, 38

  \bibitem[\protect\citeauthoryear{Adelman-McCarthy et al.}{2006b}]{adelman06b}
    Adelman-McCarthy J., et al., 2006b, in preparation

  \bibitem[\protect\citeauthoryear{Baldry et al.}{2004}]{baldry04}
    Baldry, I.K., Glazebrook, K., Brinkmann, J.,  Ivezic, Z., 
    Lupton, R.H., Nichol R.C., Szalay A.S., 2004, ApJ, 600, 681
 
  \bibitem[\protect\citeauthoryear{Bardeen et al.}{1986}]{bbks}
    Bardeen J.M., Bond J.R., Kaiser N., Szalay A.S., 1986, ApJ, 304,15
    
  \bibitem[\protect\citeauthoryear{Baugh \& Efstathiou}{1993}]{baugh93}
    Baugh C., Efstathiou G., 1993, MNRAS, 265, 145

  \bibitem[\protect\citeauthoryear{Blake et al.}{2006}]{blake06}
    Blake C., Collister A., Bridle S., Lahav O., 2006, astro-ph/0605303

  \bibitem[\protect\citeauthoryear{Blanton et al.}{2003a}]{blanton03a}
    Blanton M.R., Lin H., Lupton R.H., Maley F.M., Young N., Zehavi
    I., Loveday J., 2003a, AJ, 125, 2276

  \bibitem[\protect\citeauthoryear{Blanton et al.}{2003b}]{blanton03b}
    Blanton M.R., et al., 2003b, ApJ, 592, 819

  \bibitem[\protect\citeauthoryear{Blanton et al.}{2004}]{blanton04a}
    Blanton M.R., et al., 2004, AJ, 129, 2562

  \bibitem[\protect\citeauthoryear{Bond \& Efstathiou}{1984}]{bond84}
    Bond, J.R. \& Efstathiou, G. 1984, ApJ, 285, L45

  \bibitem[\protect\citeauthoryear{Bond \& Efstathiou}{1987}]{bond87}
    Bond, J.R., \& Efstathiou, G., 1987, MNRAS, 226, 655

  \bibitem[\protect\citeauthoryear{Cole et al.}{2005}]{cole05}
    Cole S., et al., 2005, MNRAS, 362, 505

  \bibitem[\protect\citeauthoryear{Coles \& Jones}{1991}]{coles91}
    Coles P., Jones B., 1991, MNRAS, 248, 1

  \bibitem[\protect\citeauthoryear{Colless et al.}{2001}]{colless01}
    Colless M., et al., 2001, MNRAS, 328, 1039

  \bibitem[\protect\citeauthoryear{Colless et al.}{2003}]{colless03}
    Colless M., et al., 2003, astro-ph/0306581

  \bibitem[\protect\citeauthoryear{Connolly et al.}{2002}]{connolly02}
    Connolly A., et al., 2002, ApJ, 579, 42

  \bibitem[\protect\citeauthoryear{Cooray \& Sheth}{2002}]{cooray02}
    Cooray A., Sheth, R., 2002, Physics Reports, 372, 1

  \bibitem[\protect\citeauthoryear{Dodelson et al.}{2002}]{dodelson02}
    Dodelson S., et al., 2002, ApJ, 572, 140

  \bibitem[\protect\citeauthoryear{Efstathiou et al.}{1990}]{efstathiou90}
    Efstathiou G., Sutherland, W.J., \& Maddox, S.J., 1990, 
    Nature, 348, 705

  \bibitem[\protect\citeauthoryear{Efstathiou \& Bond}{1999}]{efstathiou99}
    Efstathiou G., Bond J.R., 1999, MNRAS, 304, 75

  \bibitem[\protect\citeauthoryear{Efstathiou \& Moody}{2001}]{efstathiou01}
    Efstathiou G., Moody S.J., 2001, MNRAS, 325, 1603

  \bibitem[\protect\citeauthoryear{Eisenstein \& Hu}{1998}]{eisenstein98}
    Eisenstein D.J., Hu W., 1998, ApJ, 496, 605

  \bibitem[\protect\citeauthoryear{Eisenstein et al.}{1999}]{eisenstein99}
    Eisenstein D.J., Hu W., Tegmark M., 1999, ApJ, 518, 2

  \bibitem[\protect\citeauthoryear{Eisenstein et al.}{2001}]{eisenstein01}
    Eisenstein, D.J., Annis, J., Gunn, J.E., Szalay, A.S., 
    Connolly, A.J., Nichol, R.C., et al., 2001, AJ, 122, 2267
    
  \bibitem[\protect\citeauthoryear{Eisenstein et al.}{2005}]{eisenstein05}
    Eisenstein D.J., et al., 2005, ApJ, 633, 560

  \bibitem[\protect\citeauthoryear{Eisenstein et al.}{2006}]{eisenstein06}
    Eisenstein D.J., Seo H.-J., White M., 2006, astro-ph/0604361

  \bibitem[\protect\citeauthoryear{Feldman et al.}{1994}]{FKP}
    Feldman H.A., Kaiser N., Peacock J.A., 1994, MNRAS, 426, 23

  \bibitem[\protect\citeauthoryear{Fukugita et al.}{1996}]{fukugita96}
    Fukugita M., Ichikawa T., Gunn J.E., Doi M., Shimasaku K.,
    Schneider D.P., 1996, AJ, 111, 1748

  \bibitem[\protect\citeauthoryear{G\'orski et al.}{2005}]{gorski05}
    G\'orski, K.M., Hivon E., Banday A.J., Wandelt B.D., Hansen F.K., 
    Reinecke M., Bartelmann M., 2005, ApJ, 622, 759

  \bibitem[\protect\citeauthoryear{Gott et al.}{2005}]{gott05}
    Gott J.R., et al., 2005, ApJ, 624, 463

  \bibitem[\protect\citeauthoryear{Gunn et al.}{1998}]{gunn98} 
    Gunn J.E., et al., 1998, AJ, 116, 3040

  \bibitem[\protect\citeauthoryear{Gunn et al.}{2006}]{gunn06} 
    Gunn J.E., et al., 2006, AJ, 131, 2332

  \bibitem[\protect\citeauthoryear{Hinshaw et al.}{2006}]{hinshaw06} 
    Hinshaw G., et al., 2006, ApJS submitted, astro-ph/0603451

  \bibitem[\protect\citeauthoryear{Hogg et al.}{2001}]{hogg01} 
    Hogg D.W., Finkbeiner D.P., Schlegel D.J., Gunn J.E., 2001, AJ,
    122, 2129

  \bibitem[\protect\citeauthoryear{Holtzman}{1989}]{holtzman89}
    Holtzman J.A. 1989, ApJS, 71,1

  \bibitem[\protect\citeauthoryear{Huetsi}{2006}]{huetsi06} 
    Huetsi G., 2006, A\&A, 449, 891

  \bibitem[\protect\citeauthoryear{Ivezic et al.}{2004}]{ivezic04}
    Ivezic Z., et al., 2004, AN, 325, 583

  \bibitem[\protect\citeauthoryear{Kaiser}{1987}]{kaiser87}
    Kaiser N. 1987, MNRAS, 227, 1

  \bibitem[\protect\citeauthoryear{Lupton et al.}{1999}]{lupton99}
    Lupton R.H., Gunn J.E., Szalay A.S., 1999, AJ, 118, 1406

  \bibitem[\protect\citeauthoryear{Lupton et al.}{2001}]{lupton01}
    Lupton R., Gunn J.E., Ivezic Z., Knapp G.R., Kent S., Yasuda N.,
    2001, in ASP Conf. Ser. 238, Astronomical Data Analysis Software
    and Systems X, ed. F.R. Harnden Jr, F.A. Primini, H.E. Payne (San
    Francisco Astr. Spc. Pac.); astro-ph/0101420

  \bibitem[\protect\citeauthoryear{Maddox et al.}{1990}]{maddox90} 
    Maddox S.J., Efstathiou G., Sutherland W.J., Loveday J., 
    1990, MNRAS, 243, 692

  \bibitem[\protect\citeauthoryear{Maddox et al.}{1996}]{maddox96} 
    Maddox S.J., Efstathiou G., Sutherland W.J., 1996, MNRAS, 283, 1227
 
  \bibitem[\protect\citeauthoryear{Norberg et al.}{2001}]{norberg01}
    Norberg P., et al., 2001, MNRAS, 328, 64

  \bibitem[\protect\citeauthoryear{Norberg et al.}{2002}]{norberg02}
    Norberg P., et al., 2002, MNRAS, 332, 827

  \bibitem[\protect\citeauthoryear{Padmanabhan et al.}{2006}]{padmanabhan06}
    Padmanabhan N., et al., 2006, ApJ submitted, astro-ph/0605302

  \bibitem[\protect\citeauthoryear{Padilla \& Baugh}{2003}]{PB03}
    Padilla N.D., Baugh C.M., 2003, MNRAS, 343, 796

  \bibitem[\protect\citeauthoryear{Page et al.}{2006}]{page06} 
    Page L., et al., 2006, ApJS submitted, astro-ph/0603450

  \bibitem[\protect\citeauthoryear{Park et al.}{1994}]{park94} 
    Park C., Vogeley M.S., Geller M.J., Huchra J.P., 1994, ApJ, 431,
    569

  \bibitem[\protect\citeauthoryear{Peacock \& Smith}{2000}]{peacock00}
    Peacock J.A., Smith R.E., 2000, MNRAS, 318, 1144

  \bibitem[\protect\citeauthoryear{Peebles \& Yu}{1970}]{peebles70} 
    Peebles, P. J. E. \& Yu, J. T. 1970, ApJ, 162, 815

  \bibitem[\protect\citeauthoryear{Percival et al.}{2001}]{percival01}
    Percival W.J., et al., 2001, MNRAS, 327, 1297

  \bibitem[\protect\citeauthoryear{Percival et al.}{2002}]{percival02}
    Percival W.J., et al., 2002, MNRAS, 337, 1068

  \bibitem[\protect\citeauthoryear{Percival et al.}{2004}]{PVP}
    Percival W.J., Verde L., Peacock J.A., 2004, MNRAS, 347, 645

  \bibitem[\protect\citeauthoryear{Percival \& Brown}{2006}]{percival06a}
    Percival W.J., Brown M.L., 2006, MNRAS accepted, astro-ph/0604547

   \bibitem[\protect\citeauthoryear{Percival et al.}{2006}]{percival06b}
    Percival W.J., et al., 2006, ApJL submitted, astro-ph/0608635

  \bibitem[\protect\citeauthoryear{Pier et al.}{2003}]{pier03}
    Pier J.R., et al., 2003, AJ, 125, 1559

  \bibitem[\protect\citeauthoryear{Pope et al.}{2004}]{pope04}
    Pope A.C., et al., 2004, ApJ, 607, 655

  \bibitem[\protect\citeauthoryear{Press et al.}{1992}]{press92}
    Press W.H., Teukolsky S,A., Vetterling W.T., Flannery B.P.,
    1992, Numerical recipes in C. The art of scientific computing, 
    Second edition, Cambridge: University Press. 

  \bibitem[\protect\citeauthoryear{Saunders et al.}{2000}]{saunders00}
    Saunders W., et al., 2000, MNRAS, 317, 55

  \bibitem[\protect\citeauthoryear{Schlegel et al.}{1998}]{schlegel98}
    Schlegel D.J., Finkbeiner D.P., Davis M., 1998, ApJ, 500, 525

  \bibitem[\protect\citeauthoryear{Scoccimarro et al.}{2006}]{scoccimarro06}
    Scoccimarro R., et al., 2006, in prep.

  \bibitem[\protect\citeauthoryear{Scranton et al.}{2002}]{scranton02}
    Scranton R., et al. 2002, ApJ, 579, 48

  \bibitem[\protect\citeauthoryear{Seljak}{2000}]{seljak00}
    Seljak U., 2000, MNRAS, 318, 203
    
  \bibitem[\protect\citeauthoryear{Seljak}{2001}]{seljak01}
    Seljak U., 2001, MNRAS 325, 1359

  \bibitem[\protect\citeauthoryear{Shectman et al.}{1996}]{shectman96}
    Shectman S.A., Landy S.D., Oemler A., Tcuker D.L., Lin H.,
    Kirshner R.P., Schechter P.L., 1996, ApJ, 470, 172

  \bibitem[\protect\citeauthoryear{Silk}{1968}]{silk68}
    Silk J., 1968, ApJ, 151, 459

  \bibitem[\protect\citeauthoryear{Smith et al.}{2002}]{smith02}
    Smith J.A., 2002, AJ, 123, 2121

  \bibitem[\protect\citeauthoryear{Smith et al.}{2003}]{smith03}
    Smith R.E., et al., 2003, MNRAS, 341, 1311

  \bibitem[\protect\citeauthoryear{Spergel et al.}{2003}]{spergel03}
    Spergel D.N, et al., 2003, ApJS, 148, 175

  \bibitem[\protect\citeauthoryear{Spergel et al.}{2006}]{spergel06}
    Spergel D.N, et al., 2006, ApJS submitted, astro-ph/0603449

  \bibitem[\protect\citeauthoryear{Stoughton et al.}{2002}]{stoughton02}
    Stoughton C., et al., 2002, AJ, 123, 485

  \bibitem[\protect\citeauthoryear{Strateva et al.}{2001}]{strateva01}
    Strateva, I., et al., 2001, AJ, 122, 1861

  \bibitem[\protect\citeauthoryear{Strauss et al.}{2002}]{strauss02}
    Strauss M.A., et al., 2002, AJ, 124, 1810

  \bibitem[\protect\citeauthoryear{Sunyaev \& Zel'dovich}{1970}]{sunyaev70}
    Sunyaev, R.A., \& Zel'dovich, Ya.B., 1970,  Astrophys. \& Space Science, 
    7, 3

   \bibitem[\protect\citeauthoryear{Szalay et al.}{2003}]{szalay03}
    Szalay A., et al., 2003, ApJ, 591, 1

  \bibitem[\protect\citeauthoryear{Tadros et al.}{1999}]{tadros99}
    Tadros H., et al., 1999, MNRAS, 305, 527

  \bibitem[\protect\citeauthoryear{Tegmark et al.}{1997}]{tegmark97} 
    Tegmark M., Taylor A.N., Heavens A.F., 1997, ApJ, 480, 22

   \bibitem[\protect\citeauthoryear{Tegmark et al.}{2002}]{tegmark02}
    Tegmark M., et al., 2002, ApJ, 571, 191

   \bibitem[\protect\citeauthoryear{Tegmark et al.}{2004}]{tegmark04}
    Tegmark M., et al., 2004, ApJ, 606, 702

   \bibitem[\protect\citeauthoryear{Tegmark et al.}{2006}]{tegmark06}
    Tegmark M., et al., 2006, PRD submitted, astro-ph/0608632

   \bibitem[\protect\citeauthoryear{Tucker et al.}{2006}]{tucker06}
    Tucker D., et al., 2006, AJ, in press

  \bibitem[\protect\citeauthoryear{Yahata et al.}{2006}]{yahata06}
    Yahata K., et al., 2006, PASP submitted, astro-ph/0607098

  \bibitem[\protect\citeauthoryear{Vogeley et al.}{1992}]{vogeley92}
    Vogeley M.S., Park C., Geller M.J., Huchra, J.P., 1992, ApJ, 391, L5

  \bibitem[\protect\citeauthoryear{Vogeley \& Szalay}{1996}]{vogeley96}
    Vogeley M.S., Szalay A.S., 1996, ApJ, 465, 34

  \bibitem[\protect\citeauthoryear{York et al.}{2000}]{york00}
    York, D.G., et~al., 2000, AJ, 120, 1579

  \bibitem[\protect\citeauthoryear{Zehavi et al.}{2002}]{zehavi02}
    Zehavi I., et al., 2002, ApJ, 571, 172

  \bibitem[\protect\citeauthoryear{Zehavi et al.}{2005a}]{zehavi05a}
    Zehavi I., et al., 2005a, ApJ, 621, 22

  \bibitem[\protect\citeauthoryear{Zehavi et al.}{2005b}]{zehavi05b}
    Zehavi I., et al., 2005b, ApJ, 630, 1

\end{thebibliography}
\end{document}